\documentclass[letterpaper,aps,showpacs,showkeys,eqsecnum]{revtex4} 
 
\usepackage{slashed} 
\usepackage{anysize} 
\marginsize{2cm}{2cm}{1cm}{1cm} 
\usepackage{bm} 
\usepackage{amsfonts} 
\usepackage{amsmath} 
\usepackage{amssymb} 
\usepackage{graphicx} 
\usepackage{epstopdf} 
 
\newcommand{\be}{\begin{eqnarray}}
\newcommand{\ee}{\end{eqnarray}}

\begin{document}

\title{ Proton's Electromagnetic Form Factors from a Non-Power Confinement Potential  }

\author{ M.\ Kirchbach}\email{mariana@ifisica.uaslp.mx}  

\affiliation{Instituto de F{\'{i}}sica, Universidad Auto\'noma de San Luis Potos{\'{i}},
Av. Manuel Nava 6, Zona Universitaria,
San Luis Potos{\'{i}}, S.L.P. 78290, M\'exico}
\author{C.B.Compean}
\affiliation{Facultad de Ciencias, Universidad Aut\'onoma de San Luis Potos{\'{i}}
Lateral Av. Salvador Nava s/n,
San Luis Potos{\'{i}}, S.L.P. 78290, M\'exico
}

\begin{abstract}
The electric-charge, and magnetic-dipole form factors of the proton are calculated from an underlying constituent quark picture of hadron structure based on a potential shaped after a cotangent function, which has the properties of being both conformally symmetric and color confining,
finding adequate  reproduction of a variety of related data.

\end{abstract}


\maketitle

\tableofcontents

\section{Introduction}
\label{one}
Constituent quark model descriptions of hadron properties, such as excitation spectra, decay modes, or electromagnetic form-factors,  employ quantum few-body problems techniques based on effective potentials \cite{Rayzuddin} supposed to capture to some extent the essentials of the fundamental confining 
strong interaction.
The potentials of widest spread in the literature are shaped after power-functions of the relative distances between the quarks,
and among them one encounters for example (i) the  infinite-power square-well potential, $V_{SW}(r)=1/r^\infty =0,\quad  0\leq r\leq r_0$, and $V(r)=r^\infty=\infty $ for, $ -\infty <r<0$ and  $ r>r_0$, 
(ii) the harmonic oscillator, $V_{HO}(r)=\omega^2 r^2/2$, (iii) the Cornell potential, $V_{C}(r)=-\alpha/r +\beta r$, etc.
Such models usually require a significant number of free parameters to produce wave functions of the quark systems capable to account for the specific  of a variety of   data compiled in \cite{PART}.  One of the reasons 
for this circumstance can be  related to the mismatch between the symmetry properties of the power-potentials and the symmetries of the fundamental 
strong interaction like the conformal symmetry, which manifests itself among others by the walking of the strong coupling $\alpha_s$ to a fixed value in the infrared regime of QCD \cite{Andre} and  the notable  hydrogen-like degeneracies appearing in the mass distributions of the unflavored mesons, a phenomenon addressed  for example  in  \cite{Afonin1}, \cite{Afonin3}, \cite{Afonin2}. In order to improve this aspect of the quark models, it naturally comes to ones mind to explore  more complicated potential functions. The observation that the infinite square well has same spectrum  as the $\left[2\csc^2(\pi r/r_0)-1\right]$ potential, 
referred to in  \cite{Khare} as the ``super-symmetric partner'' to $ V_{SW}(r/r_0)$,  seems to point toward  the exactly solvable  trigonometric potentials known from the super-symmetric quantum mechanics (SUSY-QM), as possible upgrades to the power-potentials. 
Indeed, several of the finite power-potential series in use can be viewed  as first terms  in the infinite series expansions of properly designed trigonometric functions. Specifically, the (here dimensionless) inverse square distance term, $R^2/r^2$ where $R$ is a matching length parameter,  approximates $\csc^2 (r/R)$ at small $(r/R)$ values. The linear plus harmonic oscillator potential can be viewed as an  approximation to  the trigonometric Scarf potential, 
\begin{eqnarray}
V_{tSc}\left(\frac{r}{R}\right)&=&[b^2+a(a+1)]\sec^2\left(\frac{r}{R}\right) -b(2a+1)\sec\, \left(\frac{r}{R} \right)\tan\, 
\left( \frac{r}{R}\right)\nonumber\\
&\approx& -b(2a+1)\frac{r}{R} +[b^2+a(a+1)]\left(\frac{r}{R}\right)^2,
\label{GL01}
\end{eqnarray}
by the first terms of its series expansion,  
while the Cornell potential  could be viewed as a truncation of the series expansion of the  cotangent function according to, 
\begin{equation}
-b\cot\left(\frac{r}{R}\right) \approx -b\frac{R}{r}   +\frac{b}{3}\frac{r}{R}.
\label{Gl02}
\end{equation} 
The principal advantage of trigonometric- over  finite power potentials  lies not that much in the exact solubility of the former, but 
rather in their symmetry properties, which  show up in certain appropriately chosen variables.   For example,  while  the centrifugal barrier, $\ell(\ell+1)/r^2$,  and the Cornell potential are only rotationally symmetric, their trigonometric extensions towards  $\ell(\ell+1)\csc^2(r/R)$ and $-\cot (r/R)$ have the  higher  $O(4)$ symmetry, just as would be required by the conformal symmetry at the level of the excitations.
This is visible from the fact that the  stationary Schr\"odinger  wave equation, describing (upon separation of center-of mass and relative, $r/R$,  coordinates) the one-dimensional radial part

\begin{eqnarray}
\left[
-\frac{\hbar^2c^2}{R^2}
\frac{{\mathrm d}^2}{{\mathrm d}\chi^2}
+\frac{\hbar^2c^2}{R^2}
\frac{\ell (\ell +1)}{\sin^2\chi } -2\frac{\hbar^2c^2b^2}{R^2}\cot \chi\right]
U_{n\ell} (\chi)&=&{\mathcal E}^2 U_{n\ell} (\chi),
\,\, \chi= \frac{r}{R}\in \left[0,\pi\right],
\label{Glch1}
\end{eqnarray}
of a  two-body  wave function,  with $n$ being  the node-number, and $\ell $ the relative angular momentum value, can be transformed through  the change 

\begin{equation}
U_{n\ell} (\chi)Y_\ell^m(\theta,\varphi)=\frac{\Psi_{n\ell}  (\chi)}{\sin\chi}Y_\ell^m(\theta,\varphi)=\Psi^{tot}_{n\ell}(\chi,\theta,\varphi),
\label{Gl3}
\end{equation}
to quantum motion on the three dimensional hypersphere, $S^3$, according to
\begin{eqnarray}
\frac
{\hbar^2c^2}{R^2}
\left[{\mathcal K}^2 (\chi,\theta,\varphi)-
2b^2\cot \chi\right] \Psi^{tot}_{n\ell}(\chi,\theta,\varphi)&=&
\left({\mathcal E}^2 -\frac{\hbar^2c^2}{R^2}\right)  \Psi^{tot}_{n\ell} (\chi,\theta,\varphi),\nonumber\\
{\mathcal E}^2= \frac{\hbar^2c^2}{R^2}(K+1)^2 -\frac{\hbar^2c^2}{R^2}\frac{b^2}{(K+1)^2}, \quad K=n+\ell.
\label{Gl4}
\end{eqnarray}
Here, $\chi$ and $R$ take in their turn the r\'ole of the second polar angle, and hyper-spherical radius, $\theta$ and $\varphi$ are the polar and azimuthal angels in ordinary three space, ${\mathcal K}^2(\chi,\theta,\varphi)$ stands for the squared four-dimensional angular momentum operator, and $K$ for its value. 
In other words, ${\mathcal K}^2(\chi,\theta,\varphi)$ represents  
the  angular part of the Laplacian $\Delta_4(R,\chi,\theta,\varphi)$ of the four dimensional Euclidean space in global coordinates according to,
\begin{eqnarray}
\Delta_4(R,\chi,\theta,\varphi)&=&\frac{1}{R^3}\frac{\partial}{\partial R} R^3\frac{\partial}{\partial R}-\frac{1}{R^2}{\mathcal K}^2(\chi,\theta,\varphi),
\label{E4_Lplc}
\end{eqnarray}
meaning that at a constant hyper-radius, $R=$const, one finds it expressed as, 
\begin{eqnarray}
\frac{1}{R^2}{\mathcal K}^2(\chi,\theta,\varphi) =-\Delta_4(R,\chi,\theta, \varphi)|_{R=const}= \frac{1}{R^2\sin^2\chi}
\frac{\partial}{\partial\chi}\sin^2\chi\frac{\partial}{\partial \chi} +\frac{L^2(\theta,\varphi) }{R^2\sin^2\chi}, 
\label{Gl5}
\end{eqnarray}
where $L^2(\theta,\varphi)$ is the ordinary operator of the squared angular momentum whose eigenfunctions are the spherical harmonics, $Y_\ell^m(\theta,\varphi)$. In this way, the $\csc^2\chi $ term acquires meaning of  the centrifugal term on $S^3$. 
Now, the product of $U_{n\ell} (\chi)$ by the standard spherical harmonic, $Y_\ell^m(\theta,\varphi)$ in (\ref{Gl3}) represents the  complete three dimensional ``curved''  wave function, $\Psi^{tot}_{n\ell} (\chi, \theta, \varphi)$. \\

\noindent

There is one particularly remarkable aspect of the equations (\ref{E4_Lplc})-(\ref{Gl5}) which is that under the variable changes,
\begin{equation}
R=e^\tau, \quad R\sin\chi=r,
\label{varchng}
\end{equation} 
with $r$ being the  radial variable on the equatorial disc of $S^3$, a plane three dimensional Euclidean space, 
the Laplacian $\Delta_4(R,\chi,\theta,\varphi)$ is transformed to 
\begin{equation}
\Delta_4(R,\chi,\theta,\varphi)\longrightarrow e^{-2\tau}\left[ \frac{\partial ^2}{\partial \tau^2} - \frac{1}{r^2}\frac{\partial }{\partial r}r^2\frac{\partial }{\partial r} -\frac{1}{r^2}\frac{L_2^2(\varphi)}{\sin^2\theta }\right],
\label{effctv_Mnk}
\end{equation}
in the approximation $R\sin\chi \approx R\chi$. Here, $L^2_2(\varphi)$ is the operator of the squared angular momentum on the plane (for details see \cite{Fubini}).
The latter equation is the Laplacian of an ``effective''  Minkowskian space-time with the place of the ordinary time variable being occupied by the so called ``conformal time'' $\tau$, the logarithm of the hyper radius. In this way, a Minkowskian metric can be associated with the closed space. This is an essential point to which we shall come back in due place below.

\noindent 
Back to (\ref{Gl4}), all states with $n+\ell=K$ and $K$ integer (the value of the four-dimensional angular momentum) have same energy. 
Along these lines  the excitations of the unflavored mesons over the ground state energy (treated as a parameter) could be satisfactorily adjusted in \cite{EPJA16}, \cite{Addendum_EPJA} up to about 2500 MeV  by means of the constants $b$, and $R$. Moreover, there  we also presented along the lines of refs.~\cite{Belitsky}, \cite{Gorsky}  rigorous mathematical considerations relating the cotangent function to a cusped  Wilson loop on $S^3$, which allowed in due course to reveal its  color confining nature. Namely, we showed that the magnitude of the cotangent function can be expressed in terms of the strong coupling $\alpha_s$ and the number of colors, $N_c$, as
\begin{eqnarray}
2b=\alpha_sN_c.
\label{Gl2}
\end{eqnarray}
In effect, in the coordinates of (\ref{Gl5}),  the conformal and color confining properties of the cotangent function have been made manifest.\\

\noindent 
We then inserted  the  parametrization given by (\ref{Gl2})  in (\ref{Gl4}) to perform our data analyzes, and interpreted $\alpha_s$ as an effective QCD inspired  potential parameter.   
Upon extracting the  $\alpha_s$ values from the mass distribution of 71 measured mesons, organized into four 
Regge-trajectory families,  we found that they quite conveniently matched data for the mass ranges of the ground state mesons  under consideration.
Encouraged by the satisfactory meson data analyzes by the potential in (\ref{Glch1}), we here aim to extend the method briefly reviewed above to fermion description with the task to test the proton's electromagnetic form factors.

The article is structured as follows. The next section is dedicated to the Dirac equation gauged by the cotangent potential. There, we consider this equation in a specific approximation which then allows for exact solubility  by the aids of techniques established by  the super-symmetric quantum mechanics (SUSY-QM), and construct the Dirac spinors. In due course we encounter that the upper and lower Dirac spinor components satisfy  a pair of coupled one-dimensional  Schr\"odinger equations 
which parallel in the infrared  the two coupled one-dimensional  equations of the Light-Front Holographic QCD \cite{Review}, \cite{BdTD}. In Section 3 we employ the Dirac spinors obtained in this way 
in the calculations of the proton's and neutron's  electric-charge,  and  magnetic dipole form-factors, as well as in the
ratio of the proton's form-factors, finding good agreement with data. The text closes with a concise Summary and  Conclusion Section.

\section{Gauging on $S^3$  the  Dirac equation with the cotangent  potential  and the road of the super-symmetric quantum mechanics to its solutions} 
\label{Dirac}
In this section we formulate the Dirac equation gauged by the cotangent potential and construct its solutions with the aim to employ them in the subsequent section in the calculation of the proton's electromagnetic form factors. Among the many coordinates in which a Dirac equation with a cotangent gauge potential can be formulated we choose the polar coordinates  corresponding to the three-dimensional space of a constant curvature given by the hypersphere $S^3$, because this specifics chart provides a stage suited for modelling the confinement phenomenon in so far as no free charges can exist on such spaces, a textbook knowledge  \cite{LandLif}. The minimal charge configurations allowed to exist on such a geometry are color dipoles, i.e. color neutral states, just as required by confinement, and the potential produced by a color dipole is just a cotangent function \cite{EPJA16},\cite{Addendum_EPJA}, a reason for which we shall frequently refer to this potential as ``color confining dipole  (CCD)'' interaction. 
Noticing furthermore  that instantaneous potentials describe virtual processes happening outside of the causal light cone, allows to interpret the dynamics in (\ref{Gl4})  as quantum motion on a hypersphere located outside the light cone. Such a geometry can appear as (the only) closed space-like  geodesic of a four dimensional hyperboloid of one sheet, a so called $dS_4$ space, known to foliate the space-like region within the framework of the so called  ``de Sitter special relativity''  \cite{Pereira}, which hypothesizes the virtual  region of the Minkowskian space  to have one more space-like dimension. More details can be found in \cite{EPJA16}.  Below  we shall incorporate the cotangent color confining dipole function as a gauge interaction in the Dirac equation. 
Using  Riemannian spaces  to simulate gauge interactions at distinct energy scales (regimes), is in principle  legitimate according to \cite{Goeckeler}, because of  the dichotomy  between Riemann's curvature and field-strength tensors, though non-Abelian theories require special care as in this case the Gauss law, among others,  becomes more involved \cite{Serna}.
It needs furthermore  to be admitted  that so far only for  $SU(2)$ non-Abelian gauge group  the identification  of gauge spaces by a 3D Riemannian manifolds (as is the $S^3$  geodesic of the $dS_4$ Riemannian space from above) could be  fully justified  in the literature, while the   $SU(3)_c$ case of QCD is  lots more complicated  \cite{Haagensen} and still under investigation. In view of this,   we here limit ourselves to  consider the cotangent function from above as a Riemannian space inspired phenomenological  QCD interaction in the infrared. \\

The free Dirac equation on $S^3$ is well elaborated in the literature and is given by
\begin{eqnarray}
\left[i\hbar c \widetilde {\nabla}_\mu{\widetilde \gamma}^\mu(x) -mc^2 \right]\psi(x)&=&0,\nonumber\\
\lbrace {\widetilde \gamma}^\mu(x), {\widetilde \gamma}^\nu(x)\rbrace &=&2g^{\mu\nu}(x),\quad x=(x_0,x_1,x_2,x_3),
\label{bessis_1}
\end{eqnarray}
where $\psi (x)$ is a four-component spinor, $g^{\mu\nu}(x)$ is the $S^3$ metric tensor, 
${\widetilde \nabla}_\mu(x)$ is a {\it spin covariant} derivative on $S^3$,
while ${\widetilde \gamma}^\mu(x)$ are the Dirac matrices on the manifold under consideration (for details see \cite{Bessis}, among others).
It is common to re-parametrize the hypersphere, defined as $x_0^2+x_1^2+x_2^2+x_3^2=R^2$, in terms of global coordinates according to, $x_0=R\cos\chi$, $x_3=R\sin\chi\cos\theta$, $x_1=R\sin\chi \sin\theta \cos \varphi$, and 
$x_2=R\sin\chi\sin\theta\sin\varphi$, with $\chi,\theta \in [0,\pi]$, and $\varphi\in [0,2\pi]$.
The algebra of (\ref{bessis_1}) has been  worked out in great detail in \cite{Bessis}, \cite{Bessis2} in connection with the particular case of a $\psi(x)$ coupled  to a cotangent potential produced by an electromagnetic source . There the authors show that  upon variable separation, $\psi(x)\to \Psi(R,\chi,\theta,\varphi)=\Psi_{j\ell} (R,\chi)Y_L^M(\theta,\varphi)$, with $Y_L^M(\theta,\varphi)$ being the standard spherical harmonics,  one encounters in the variable $\chi$, frequently termed to as  ``quasi-radial'' in the literature, the following system of two linear coupled equations ,
\begin{eqnarray}
\frac{\hbar c}{R}\left( \frac{{\mathrm d}}{{\mathrm d}\chi} +\frac{
(-1)^{j+\ell +\frac{1}{2}}\left( j+\frac{1}{2}\right)}{\sin\chi } \right)G_{j \ell }(\chi)
&=&\left(E_{j\ell }+mc^2+ \frac{\hbar c}{R}\alpha Z\cot\chi  \right)F_{j \ell }(\chi),\nonumber\\
\frac{\hbar c}{R}\left(- \frac{{\mathrm d}}{{\mathrm d}\chi} +\frac{
(-1)^{j+\ell +\frac{1}{2}}\left( j+\frac{1}{2}\right)}{\sin\chi } \right)F_{j\ell }(\chi)
&=& \left(mc^2 -E_{j\ell}+ \frac{ \hbar c}{R}\alpha Z \cot \chi \right)G_{j \ell }(\chi),\nonumber\\
 j&=&\ell \pm \frac{1}{2}, \, \ell\geq 0,
\label{Dirac_pre}
\end{eqnarray}
where $\alpha=e^2/(4\pi\hbar c) $ is the fundamental electromagnetic constant, and  $Z$ is the  number of charges in the potential source.
{}Furthermore, $G_{j\ell }(\chi)$, and $ F_{j \ell }(\chi)$, relate to  the respective upper and lower components of the ``quasi-radial'' Dirac spinor, $\Psi_{j\ell}(R,\chi)$,  according to
\begin{equation}
\Psi_{j \ell }(R,\chi)= \left(
 \begin{array}{c}
\frac{iG_{ j \ell  }(\chi)}{R\sin \chi}\\
\frac{F_{j \ell  }(\chi)}{R\sin \chi}
\end{array}
\right), \quad \mbox{with} \quad \int_0^\pi \left( G^2_{j \ell }(\chi) + F^2_{j \ell }(\chi)\right){\mathrm d}\chi =1.
\label{quasiradial_Dirac_spinor}
\end{equation}
 Our case relates to (\ref{Dirac_pre}) through the replacements, 
\begin{eqnarray}
-\frac{\hbar c}{R}\alpha Z\cot\chi&\Rightarrow&-\frac{\hbar c}{R}\alpha_s N_c\cot \lambda \chi, \quad \alpha_s=\frac{g_s^2}{4\pi \hbar c}, \quad \lambda\chi \in [0,\pi].
\label{our_gpt}
\end{eqnarray}
 The $mc^2$ parameter entering the eqs.~(\ref{Dirac_pre}) is the reduced  mass of the constituents of the  two-body spin-$1/2$ composite system, considered upon the reduction of the two-body to one-body problem to
 move in  the gauge potential. In the flat space case of the H atom  this degree of freedom is a spin-1/2 electron of a single charge. However, on $S^3$, where no single charges can be defined, 
a  spin-1/2 degree of freedom  has to be a charge dipole, such as  a spin-$1/2$ (quark)-(anti-symmetric scalar diquark) configuration of the proton. The potential 
source can be then thought of as an effective  scalar gluon-anti-gluon background. In the following, the influence of the former dipole on the latter. i.e. the tensor force between the two dipoles, will be neglected. Within this picture, $mc^2$ denotes the reduced mass of the quarkish and gluonic color dipoles.
The introduction of the $\lambda$ constant in the ``quasi-radial'' variable will become clear in due course.

The equations in (\ref{Dirac_pre}) are claimed in \cite{Ovsyuk} to be exactly solvable in terms of Heun's polynomials. 
We here instead adopt the approximation of  the kinetic term used in  \cite{Bessis},  
 \begin{equation}
\frac{\hbar c}{R\sin\lambda \chi}  \approx  \frac{\hbar c }{R}\cot \lambda \chi +{\mathcal O}\left(\frac{1}{R^2}\right),
\label{first_order_inverse_R}
\end{equation}
amounting to the following, also exactly solvable, matrix equation,
\begin{eqnarray}
\frac{\hbar c}{R}\left( 
\begin{array}{cc}
\frac{{\mathrm d}G_{j\ell }(\chi)}{{\mathrm d}\chi}& 0\\
0& \frac{{\mathrm d}F _{j \ell }(\chi)}{{\mathrm d}\chi}
\end{array}
\right)
 &+&\frac{\hbar c}{R}\cot\lambda \chi \left( 
\begin{array}{cc}
k&-\gamma\\
\gamma& -k
\end{array}
\right)
\left(
 \begin{array}{c}
G_{j \ell }(\chi)\\
F_{j \ell }(\chi)
\end{array}
\right)\nonumber\\
&=&
\left( 
\begin{array}{cc}
0&E_{j\ell }+mc^2\\
mc^2-E_{j\ell }&0
\end{array}
\right)\left(
 \begin{array}{c}
G_{j\ell }(\chi)\\
F_{j \ell }(\chi)
\end{array}
\right).
\label{Dirac_1}
\end{eqnarray}

The following notions have been introduced
\begin{equation}
k=\left( -1\right)^{\ell +j +\frac{1}{2}}\left(j+\frac{1}{2} \right), \quad \gamma=\alpha_sN_c.
\label{constants}
\end{equation}

\begin{figure}
\includegraphics[width=7.5cm]{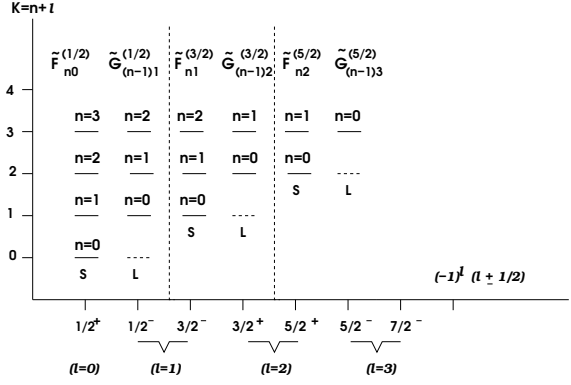}
\caption{Schematic presentation of the solutions ${\widetilde G}_{n'\ell'}^j(\chi)$ and ${\widetilde F}_{n\ell}^j(\chi)$ to the eqs.~(\ref{Pot_H1}) and (\ref{H2}), acting as the  respective upper/large (L) and lower/small  (S) components of the intermediate auxiliary spinor ${\widetilde \Psi}_{j\ell}(\chi)$ in (\ref{trnsfrmd_wafuquasiradial_Dirac_spinor}) within the  SUSY-QM scheme. On the figure we separated the spinors of different $j$  by vertical dashed lines. To be specific,  the first, second, third and etc. columns illustrate  the respective ${\widetilde \Psi}_{(1/2),0}(\chi)$, ${\widetilde \Psi}_{(3/2)1 }(\chi)$, ${\widetilde \Psi}_{(5/2) 2 }(\chi)$, and etc. spinors with rising $n$. The nomenclature used in the   ${\widetilde \Psi}_{j\ell}(\chi)$ labeling is such that the index $\ell$ belongs to the angular momentum underlying the lower component.
The states of the  highest spin possible for a given $K$  correspond to $n=0$, their $\ell$ takes the maximal  $\ell_{\mbox{ max}}=K$ value, and their spins are $j=\left(K +\frac{1}{2}\right)$. Their spinors   have all only lower components. The upper components to $n=0$ are absent  (denoted by horizontal dashed  segments) because they would require  according to (\ref{Gen_Dir_Spnr}) a node number lower by one unit than $n$, i.e. a prohibited negative value. These spinors describe  the states of the lowest energies (ground states) in a column.  In consequence,  the genuine Dirac ground state spinors, $\Psi_{j \ell}(R,\chi)$, in (\ref{quasiradial_Dirac_spinor}) (upon accounting for (\ref{trnsfrmd_wafuquasiradial_Dirac_spinor})),  have upper  and lower components of equal functional forms, distinct by a constant.
The above lying spinors are associated with solutions to the Dirac equation corresponding to  energies higher than  the ground state value, and their 
upper and lower components are of different functional forms. \label{SUSY_SPCTR}}
\end{figure}

{}From a purely technical point of view, the approximation to the Dirac equation on $S^3$ by  (\ref{Dirac_1}) allows for a treatment in exact parallel to the flat-space Dirac equation  with the Coulomb potential, and  along the line  presented, for example, in \cite{Sukumar}.
{}For this purpose the equation (\ref{Dirac_1}) has to be similarity transformed by the following matrix,
\begin{eqnarray}
D&=&\left(
\begin{array}{cc}
k+s&-\gamma\\
-\gamma&k+s
\end{array} 
\right), \quad
\quad s=\sqrt{k^2 -\gamma^2} = \sqrt{\left( j+\frac{1}{2} \right)^2 -\alpha_s^2N_c^2} ,
\label{parametri}\\
D^{-1}&=&\frac{1}{2s(s+k)}\left( 
\begin{array}{cc}
k+s&\gamma\\
\gamma& k+s
\end{array}
\right),\quad
 D\left( \begin{array}{cc}
k&-\gamma\\
\gamma&-k
\end{array}\right)D^{-1}
=\left( \begin{array}{cc}s&0\\
0&-s\end{array}\right),
\label{D_matrix}
\end{eqnarray}   
with the result being the following two coupled  linear equations

\begin{eqnarray}
\left[ \frac{\hbar c}{R} \frac{{\mathrm d}}{{\mathrm d} \chi} + W(\chi) \right]{\widetilde G}^j_{n' \ell' }(\chi)       
&=&\left(mc^2+\frac{k}{s}E_{j\ell}\right){\widetilde F}^j_{n \ell }(\chi),\label{first}\\
\left[ - \frac{\hbar c}{R}\frac{{\mathrm d}}{{\mathrm d} \chi}+ W(\chi) \right]{\widetilde F}^j_{n \ell }(\chi)&=&
\left(mc^2- \frac{k}{s}E_{j\ell} \right)  {\widetilde G}^j_{n' \ell' }(\chi),      
\label{Trnsfrmd_Dirac}\\
 \quad n'+\ell'=n+\ell, \quad W(\chi)&=&-\frac{\hbar c}{R}s\cot\lambda \chi -\frac{\gamma }{s}E_{j\ell },\label{sprPT}
\end{eqnarray}
with
\begin{eqnarray}
{\widetilde \Psi}_{j\ell }(R, \chi)=\left( 
\begin{array}{c}
 {\widetilde G}^j_{n' \ell' }(\chi)\\
{\widetilde F}^j_{n \ell }(\chi)
\end{array}
\right)&=&\left( \begin{array}{cc}
k+s&-\gamma\\
-\gamma&k+s
\end{array}\right)\left( \begin{array}{c}
 G_{ j \ell }(\chi)\\
 F_{j \ell }(\chi)
\end{array}
\right),
\label{trnsfrmd_wafuquasiradial_Dirac_spinor}
\end{eqnarray}
where we introduced  the intermediate auxiliary spinor ${\widetilde \Psi}_{j\ell}(R,\chi)$.
The inclusion of the node number $n$ into the labeling of the components ${\widetilde G}$, and ${\widetilde F}$,  of the auxiliary spinors will become clear in due course, and is advantageous 
within the SUSY-QM framework \cite{Khare}, on which \cite{Sukumar} is based. 
There, the left hand sides of the equations (\ref{first}) and (\ref{Trnsfrmd_Dirac}) are defined by the so called up- and down ladder operators, respectively, denoted by
\begin{eqnarray}
A^+(\chi)&=& \frac{\hbar c}{R} \frac{{\mathrm d}}{{\mathrm d} \chi} +W(\chi),\label{Aplus}\\
A^-(\chi)&=& -\frac{\hbar c}{R} \frac{{\mathrm d}}{{\mathrm d} \chi} +W(\chi),
\label{Aminus}
\end{eqnarray}
with the $W(\chi)$ function being termed to as a  ``super-potential''.
As it will become clear below, one  of the tasks of the  ladder operators is to shift the node number by one unit, so that $n'=n-1$.  
By means of them, the equations (\ref{first})-(\ref{Trnsfrmd_Dirac}) are cast in the elegant form of
\begin{eqnarray}
A^+(\chi){\widetilde G}^j_{n' \ell' }(\chi)       
&=&E_{j\ell}^{(1)}{\widetilde F}^j_{n \ell }(\chi), \quad n'=n-1, \quad \ell'=\ell +1,\label{SUSY1}\\
A^-(\chi) {\widetilde F}^j_{n \ell }(\chi)       
&=&E_{j\ell }^{(2)}{\widetilde G}^j_{n' \ell' }(\chi), 
\label{SUSY2}\\
E_{j\ell }^{(1)}&=&mc^2+\frac{k}{s}E_{j\ell }, \quad E_{j\ell }^{(2)}= mc^2-\frac{k}{s}E_{j\ell }, \nonumber\\
 E_{j\ell }^{(1)}&\not=&E_{j\ell }^{(2)} \quad\mbox{\footnotesize for}\quad mc^2\not=0.
\label{SUSY3}
\end{eqnarray}
It is straightforward to check that the two  coupled linear equations in (\ref{first})-(\ref{Trnsfrmd_Dirac}) are equivalent to the following two decoupled quadratic equations
\begin{eqnarray}
{\Big(}-\frac{\hbar^2c^2}{R^2}\frac{{\mathrm d}^2}{{\mathrm d}\chi^2} +V_1(\chi){\Big)}{\widetilde G}^j_{n' \ell' }(\chi)
&=&\left( \frac{k^2E_{j\ell}^2}{s^2} -m^2c^4\right){\widetilde G}^j_{n' \ell' }(\chi),
\label{Pot_H1}\\
{\Big(}-\frac{\hbar^2c^2}{R^2}\frac{{\mathrm d}^2}{{\mathrm d}\chi^2} +V_2(\chi){\Big)}
{\widetilde F}^j_{n \ell }(\chi)
&=&\left( \frac{k^2E_{j\ell }^2}{s^2} -m^2c^4\right){\widetilde F}^j_{n \ell }(\chi),
\label{H2}
\end{eqnarray}

with the potentials $V_1(\chi)$ and $V_2(\chi)$ being given as,
\begin{eqnarray}
V_1(\chi)=W(\chi)^2-\frac{\hbar c}{R}W'(\chi)&=&\frac{\hbar^2c^2}{R^2}\frac{(s+1)(s+1 -\lambda)}{\sin^2\lambda \chi}
 -2\frac{\hbar c}{R}\gamma E_{j\ell } \cot\lambda \chi\nonumber\\
& -&\frac{\hbar^2c^2}{R^2}s^2 +\frac{\gamma^2}{s^2}E_{j\ell }^2,
\label{H1}\\
V_2(\chi)= W(\chi)^2+\frac{\hbar c}{R}W'(\chi) &=& \frac{\hbar^2c^2}{R^2}\frac{s(s-\lambda )}{\sin^2\lambda \chi} 
-2\frac{\hbar c}{R}\gamma E_{j\ell }\cot\lambda \chi\nonumber\\
& -&\frac{\hbar^2c^2}{R^2}s^2 +\frac{\gamma^2}{s^2}E_{j\ell }^2.
\label{Pot_H2}
\end{eqnarray}
The dependence of the magnitude of the cotangent potential on the energy, $E_{j\ell}$, represents the genuine signature for the origin of the SUSY-QM equations from a Dirac equation.
Both the  $V_1(\chi)$, and $V_2(\chi)$ interactions are shaped after a function known in SUSY-QM under the name of the trigonometric Rosen-Morse potential, and the two, in being  in addition  isospectral, are termed to as  ``partner potentials''. In this particular case the isospectrality is ensured by  obtaining  $V_1(\chi)$  from $V_2(\chi)$  by the replacements, $s\longrightarrow s+1$, and $n'=(n-1)$, everywhere besides in the constants, thus keeping  $(s+n)=(s+1 +n' )$ unaltered (see Fig.~\ref{SUSY_SPCTR} for a graphical interpretation).

In the following, we shall set $\lambda =-1$, in which case  $s(s-\lambda)\csc^2 \lambda\chi \longrightarrow s(s+1)\csc^2 \chi$ for $V_2(\chi)$, while in $V_1(\chi)$ this term becomes, $(s+1)(s+1-\lambda)\csc^2\lambda \chi \longrightarrow (s+1)(s+2)\csc^2 \chi$, a choice which, contrary to $\lambda=1$,  will lead to everywhere \underline{finite} probability densities. In this sense, the rescaling of the ``quasi-radial'' variable by the $\lambda$ parameter introduced here by us can be viewed as a regularization scheme, different from the one used in the case of the H Atom to avoid the singularity of its Dirac ground state spinor \cite{Knut}.

\subsection{General form of the  ``quasi-radial'' Dirac spinors }

The (unnormalized) wave functions solving  (\ref{H2})-(\ref{Pot_H2}) are known and available in the literature. In the particular form constructed by \cite{jpa_05}, and reviewed in \cite{raposo}, they are given by
\begin{eqnarray}
{\widetilde F}^{j}_{n\ell}(\chi)&=&R^{-\beta_n} \sin ^{-\beta_n}\chi\, e^{-\frac{\alpha_n\chi}{2}}R_n^{(\alpha_n,\beta_n)}(\cot \chi),\label{gen_sltn}\\
\beta_n=-(n+s +1),&& \alpha_n=\frac{2\gamma E_{j\ell } R}{\hbar c (n+s+1)},\quad E_{j\ell}=\frac{mc^2s}{|k|},
\label{prmtrs_Rom_plnms}\\
{\widetilde G}^{j}_{n'\ell'}(\chi)|_{s\to s+1}&=&R^{-\beta_{n'}}\sin ^{-\beta_{n'}}\chi \,
e^{-\frac{\alpha_{n'}\chi }{2}}R_{n'}^{(\alpha_{n'},\beta_{n'})}(\cot\chi)|_{s\to s+1},\label{gen_sltn_1}\\
\beta_{n'}|_{s\to s+1}&=&\beta_n,\quad \alpha_{n'}|_{s\to s+1}=\alpha_n, \quad n'=n-1,
\label{prmtrs_Rom_plnms_2}
\end{eqnarray}
where $R^{(\alpha_n,\beta_n)}_n(\cot\chi)$ are the Romanovski polynomials \cite{raposo}. These polynomials can be obtained  with the aid of the Rodrigues formula
\begin{eqnarray}
R_n^{(\alpha_n, \beta_n)}(x) &=& \frac{1}{\omega^{(\alpha_n,\beta_n)}(x)}\frac{{\mathrm d}^n}{{\mathrm d}x^n}\left[\omega^{(\alpha_n,\beta_n)}(x) (1+x^2)^n\right], \quad x=\cot\chi,\label{Rdrgz}
\end{eqnarray}
and from the following weight function
\begin{equation}
\omega^{(\alpha_n,\beta_n)}(x)=(1+x^2)^{\beta_n }e^{-\alpha_n \cot^{-1}x}.
\end{equation}
In this fashion, the general (unnormalized) ``quasi-radial''  Dirac spinor in (\ref{quasiradial_Dirac_spinor})  take the form,
\begin{eqnarray}
\Psi_{j\ell}(R,\chi)&=&\frac{1}{2s(s+k)}\left( 
\begin{array}{cc}
(k+s)&\gamma\\
\gamma&(k+s)
\end{array}
\right)\left( 
\begin{array}{c}
\frac{{\widetilde G}^{j}_{n'\ell'=(\ell +1)}(\chi)|_{s\to s+1}}{R \sin \chi}\\
\frac{{\widetilde F}^{j}_{n\ell}(\chi)}{R\sin \chi}
\end{array}
\right)\nonumber\\
&=& \frac{1}{2s(s+k)}\left( 
\begin{array}{cc}
(k+s)&\gamma\\
\gamma&(k+s)
\end{array}
\right)\frac{R^{n+s+1}\sin^{n+s+1}\chi}{R\sin\chi}
\left( 
\begin{array}{c}
e^{
-\frac{\gamma E_{j{\ell}} R \chi }{\hbar c((n+s+1) } R_{n-1}^{(\alpha_{n},\beta_{n})}(\cot\chi)}\\
 e^{
-\frac{\gamma E_{j\ell} R \chi}{\hbar c(n+s+1) } R_n^{(\alpha_n,\beta_n)}(\cot\chi)}
\end{array}
\right).
\label{Gen_Dir_Spnr}\nonumber\\
\end{eqnarray}

These are the Dirac spinors which provide the basis for the relativistic description of the nucleon and its excitations.

\subsection{The ``quasi-radial'' Dirac spinor for the ground state of the  proton}
We here are primarily interested in the ground state for which $\ell =n=0$, $j=1/2$,  and $k=-1$, in which case the equations (\ref{first})-(\ref{Trnsfrmd_Dirac}) are,
\begin{eqnarray}
\left[ \frac{\hbar c}{R} \frac{{\mathrm d}}{{\mathrm d} \chi} + W(\chi)\right]{\widetilde G}^{(1/2)}_{n' 1 }(\chi)       
&=&\left(mc^2-\frac{1}{s+1}E_{(1/2)0}\right){\widetilde F}^{(1/2)}_{0 0 }(\chi),
\label{first_2}\\
\left[ -\frac{\hbar c}{R} \frac{{\mathrm d}}{{\mathrm d}\chi} +W(\chi)\right]
{\widetilde F}^{(1/2)}_{00 }(\chi)
&=&\left( -\frac{1}{s+1}E_{(1/2)0} -mc^2\right)  {\widetilde G}^{(1/2)}_{n' 1 }(\chi). 
\label{Dirac_gst}
\end{eqnarray}
As explained in the caption of Fig.~\ref{SUSY_SPCTR}, in SUSY-QM  the ground state spinor corresponds to $n=0$, and is obtained upon nullifying the right hand sides of the equations (\ref{first}),  (\ref{Trnsfrmd_Dirac}).
This means that for odd $(-1)^{j+\ell +\frac{1}{2}}=(-1)$ phases, only ${\widetilde F}^j_{0 \ell }(\chi)$ can be different from zero, while 
${\widetilde G}^j_{n' \ell' }(\chi)$ has to  identically vanish, due to the prohibited  $n'=-1$ value. 
Indeed, because the coefficient $(mc^2-E_{(1/2)0}/(s+1) )$ on the rhs in (\ref{first_2}) can be zero, fixing the $E_{(1/2)0}$ value to
$E_{(1/2)0}=mc^2/(s+1)$, this equation can have solutions for  ${\widetilde F}^{(1/2)}_{0 0}(\chi)\not=0$. On the other side, because the coefficient $(-E_{(1/2)0}/(s+1)-mc^2)$ on the rhs in (\ref{Dirac_gst}) is always \underline{different}  from zero, the condition defining ${\widetilde F}^{(1/2)}_{00}(\chi)$ as a ground state
\begin{equation}
\left[ -\frac{\hbar c}{R} \frac{{\mathrm d}}{{\mathrm d} \chi} + W(\chi)\right]{\widetilde F}^{(1/2)}_{0 0 }(\chi)=0,
\end{equation}  
can be fulfilled only if ${\widetilde G}^{(1/2)}_{n' 1 }(\chi)$ in (\ref{Dirac_gst}) identically vanishes
\begin{equation}
{\widetilde G}^{(1/2)}_{n' 1 }(\chi)=0, \quad \mbox{for}\quad E_{(1/2)0}=\frac{s+1}{|k|}mc^2, \quad |k|=|-1|=1, \quad s=\sqrt{1-\alpha_s^2N_c^2}<1.
\label{gst_energy}
\end{equation}
 
In this fashion, the Dirac spinor in (\ref{quasiradial_Dirac_spinor}) corresponding to the ground state (gst) is calculated  from 
(\ref{trnsfrmd_wafuquasiradial_Dirac_spinor}) and (\ref{D_matrix}) as
\begin{eqnarray}
\Psi^{\mbox{\footnotesize gst }}_{(1/2) 0}(R,\chi)&=& N_{(1/2)0}\left(
 \begin{array}{c}
\frac{iG_{(1/2) 0}(\chi)}{R\sin\chi}\\
\frac{F_{(1/2) 0}(\chi)}{R\sin \chi}
\end{array}
\right)= N_{(1/2)0}\frac{1}{2s(s+k)}
\left(
\begin{array}{c}
\frac{i\gamma {\widetilde F}^{(1/2)}_{0 0}(\chi)}{R\sin \chi}\\
\frac{(k+s) {\widetilde F}^{(1/2)}_{0 0}(\chi)}{R\sin \chi}
\end{array}
\right),
\label{QSRD_spinor}
\end{eqnarray}
where $N_{(1/2)0}$ is a normalization constant, while ${\widetilde F}^{(1/2)}_{0 0}(\chi)$ is given by
\begin{eqnarray}
{\widetilde F}^{(1/2)}_{0 0}(\chi)&=& R^{-\beta_0}\sin^{-\beta_0 }\chi e^{-\frac{\alpha_0}{2}\chi } , \quad
\alpha_0=\frac{2\gamma mc^2 R}{|k|  \hbar c }, \quad -\beta_0=s+1,
\label{F0}
\end{eqnarray}
where use of the equation (\ref{prmtrs_Rom_plnms}), and (\ref{gst_energy}) has been made. 
Substituting (\ref{F0}) into (\ref{QSRD_spinor})  shows  that the choice of $\lambda =-1$ ensured that the spinor's pre-factor $\sin^{-\beta_0}/\sin\chi$ is finite for $s<1$, and that the spinor basis obtained in this way is free from singularities.

\subsection{Comparison of the ``quasi-radial'' Dirac spinors to the two-component  wave functions in the Light-Front Holographic QCD}

The system of equations (\ref{Pot_H1})-(\ref{Pot_H2}) resembles  in certain  sense, to be specified below,
a similar one emerging within the Light Front Holographic QCD  (LF-HQCD) \cite{Review}, \cite{BdTD}. The latter formalism has been derived within a flat Minkowski 3+1 space time and the corresponding equations reduce to one-dimensional equations in the 
light front radial variable $\zeta$ after projection on states of fixed $J_z$ and $L_z$. This method has been concluded from 
the AdS$_5$/CFT$_4$ gauge-gravity duality and is of a field-theoretical nature. There, one encounters the following two coupled stationary Schr\"odinger equations, 
\begin{eqnarray}
H^\nu(\zeta)\Psi^{n\nu}_+(\zeta)&=&\left(-\frac{{\mathrm d}^2}{{\mathrm d}\zeta^2}
+\frac{\nu^2 -\frac{1}{4}}{\zeta^2 }+\kappa^4\zeta^2  +c_+^\nu\right)\Psi^{n\nu}_+(\zeta)=M^2\Psi^{n\nu}_+(\zeta),\nonumber\\
c^\nu_+ &=& 2\kappa^2(\nu+1),
\label{Gl1}\\
{H}^{\nu+1}(\zeta)\Psi^{n (\nu+1)}_-(\zeta)&=&\left(-\frac{{\mathrm d}^2}{{\mathrm d}\zeta^2} +\frac{(\nu+1)^2 -\frac{1}{4}}{\zeta^2 }+\kappa^4\zeta^2  +c_-^\nu \right)
\Psi^{n(\nu+1)}_-(\zeta)=M^2
\Psi^{n (\nu+1)}_-(\zeta),\nonumber\\
 c_-^\nu&=&\kappa^2\nu.
\label{Gl1_n}
\end{eqnarray}

The solutions in (\ref{Gl1})- (\ref{Gl1_n}) correspond physically to a quark with spin parallel vs. anti-parallel to the proton's helicity in the nucleon quark-diquark bound state and allow to compute  the Pauli form factor from the overlap of the $\Psi_\pm$  solutions, in which case the weights of $\Psi_\pm$  solutions are equal.

The  LF-HQCD wave functions provide  a variety of dynamical predictions,  among them  hadron spectra, distribution amplitudes,  and hadron form factors. The latter are  calculated  as first-principle hadronic matrix elements of the electromagnetic current and, in being  derived  using the Drell-Yan West formalism, are frame-independent.  Also quark counting rules at high momentum transfer are satisfied.\\

\noindent
{}From a purely algebraic point of view, and leaving aside the discussion on the different conceptual backgrounds of the two methods, $AdS_5/CFT_4$ gauge-gravity duality of the latter, versus $dS_4$ special relativity with a Dirac equation on $S^3$ of the present \cite{EPJA16}, the 
LF-HQCD equations relate to our equations (\ref{Trnsfrmd_Dirac})-(\ref{sprPT}) (and vice versa) via the following replacement of  the super-potential $W(\chi)$ in (\ref{sprPT}), 
\begin{equation}
 \left[ W(\chi)=-\frac{\hbar c}{R}s\cot\lambda \chi -\frac{\gamma}{s}E_{j\ell} \right]{\longleftrightarrow }
\left[W^{LF}(\zeta)=-\frac{\nu + \frac{1}{2}}{\zeta} +\kappa^2 \zeta\right],
\label{HO_sprpt}
\end{equation}
where  $\kappa ^2$ is an external inverse length scale. 
Similarly as the super-potential in (\ref{sprPT}) generated our conformal partner interactions $V_1(\chi)$ and $V_2(\chi)$  in
(\ref{H1}) and (\ref{Pot_H2}), also the super-potential $W^{LF}(\zeta) $ in  (\ref{HO_sprpt}) generates in light front coordinates a pair of conformal confining  interactions though of an infinite range, and  composed by  inverse distance square plus  harmonic oscillator terms, and given by
\begin{eqnarray}
V^{LF}_2(\zeta)&=&4\kappa^4\zeta^2  +\frac{\nu^2-\frac{1}{4}}{\zeta^2} -c^\nu_+, \quad  \zeta\in [0,\infty),\\
V^{LF}_1(\zeta)&=&4\kappa^4\zeta^2  +\frac{(\nu+1)^2-\frac{1}{4}}{\zeta^2} -c^\nu_-,
\end{eqnarray}
with $c_+^\nu=2\kappa^2(\nu +1)$, and $c_-^\nu=2\kappa^2\nu$.
The  $V^{LF}_1(\zeta)$ and $V^{LF}_2(\zeta)$ mass spectra  emerge in turn as
$M^2_{n \nu }=4\kappa^2n$ and  $M^2_{n' \nu'}=4\kappa^2(n'+1)$ meaning that for $n'=n-1$ one finds isospectrality. Obviously, for $n=0$, and $\nu'=\nu +1$ the vacuum, same as in our case above, remains unpaired. 
The respective solutions, $|n,\nu\rangle$ and $|n'=n-1,\nu'=\nu+1\rangle $, act as the counterparts to ours
${\widetilde F}^{j\ell}_{n a}(\chi)$ and ${\widetilde G}^{j\ell}_{n'a'}(\chi)$ from (\ref{prmtrs_Rom_plnms})-(\ref{prmtrs_Rom_plnms_2}). 
Yet, the equidistance of the harmonic oscillator excitations, absent in our case,  
allows for  a global shift downwards  by a constant of all  the $M^2_{n' \nu'}$ masses, and  thereby allows to place  the $|n=0,\nu'=\nu+1\rangle$ partner to $|n=1,\nu >$ at the same mass as the originally unpaired $|n=0,\nu\rangle$ ground state. In this way,  super-symmetric partners of equal node numbers to all the states  have been created in \cite{BdTD}, after which the Light Front spinors  acquire their final  shapes according to,
\begin{eqnarray}
\Psi^{LF}(\zeta)&=&\left(
\begin{array}{c} 
(\kappa^2\zeta^2)^{\frac{\nu}{2}+\frac{1}{4}}e^{-\frac{\kappa^2\zeta^2}{2}}L_n^\nu(\kappa^2\zeta^2)\\
(\kappa^2\zeta^2)^{\frac{\nu+1}{2}+\frac{1}{4}}e^{-\frac{\kappa^2\zeta^2}{2}}L_n^{\nu+1}(\kappa^2\zeta^2)
\end{array}
\right),
\label{LFspinor}
\end{eqnarray}
where $L_n^\nu$ stand for  Laguerre's  polynomials.  The above algebraic manipulation obliterates to some extent the parallelism between  the latter equation and our (\ref{Gen_Dir_Spnr}), where the node numbers in the upper and lower components are distinct by one unit, as is inevitably the case for all SUSY-QM potentials except the Harmonic Oscillator.

 In the following section we test the ``quasi-radial'' spinor in  (\ref{QSRD_spinor})-(\ref{F0}) in the calculation of the electric-charge and magnetic-dipole form factors of the proton.

\section{The proton electric-charge  and magnetic-dipole  form factors}
\label{formfctrs}
{}Form factors provide valuable  insights into the internal structure of the nucleon. In particular, the electric-charge $G_E^p(Q^2)$,  and magnetic-dipole, $G_M^p(Q^2)$, form factors of the proton, the subject of the current section, codify to some extent  the internal electric-charge--, and mag\-ne\-ti\-za\-tion-current distributions of this  particle. This is so because in the Breit-frame, defined at zero energy transfer,  
 ${\mathbf p}=-{\mathbf p}^\prime$, they can be defined \cite{Kelley} as   
the Fourier transforms  the corresponding electric-charge-, and mag\-ne\-ti\-za\-tion-current densities 
in position space, $\rho^p(r)$, and  $ \rho^p_{\mbox{\footnotesize mgn}}(r)$,   
\begin{eqnarray}
G_E^p (Q^2) &=& \int_0^\infty \rho^p(r) e^{i{\mathbf r}\cdot {\mathbf q}}{\mathrm d}{\mathbf r} ,\label{SachsEl}\\
G_M^p(Q^2)&=&\int_0^\infty \rho^p_{\mbox {mgn}}(r)e^{i{\mathbf r}\cdot {\mathbf q}}{\mathrm d}{\mathbf r} .\label{SachsMag}
\end{eqnarray}
where the transferred four-momentum is space-like, $-q^2=Q^2\geq 0$.
In terms of Dirac spinors and their components, one finds
\begin{eqnarray}
\rho^p(r) &=& |\Psi_{j\ell}(r)|^2= G^2_{(1/2)0}(r)+F^2_{(1/2)0}(r),\label{Dir_chrg}\\
\rho^p_{\mbox{\footnotesize mgn}}(r)&=& \bar{\Psi}_{j\ell}(r)\gamma_5\Psi_{j\ell}(r) =\frac{2G_{(1/2)0}(r)F_{(1/2)0}(r)}{Q}\mu_N\frac{ \partial}{\partial r}, \label{Dir_mgnt}
\end{eqnarray}
where  $\mu_N$ stands for Bohr's nuclear magneton, $ \mu_N=\frac{\hbar c}{M_pc^2}$.
  In effect, the form factors $G_E^p(Q^2)$ and $G_M^p(Q^2)$ can be expressed as the following integrals  \cite{Lu}-\cite{Luu},
\begin{eqnarray}
G_E^p (Q^2) & = & \int_0^\infty \left(G^2_{(1/2)0}(r)+F^2_{(1/2)0}(r) \right) j_0(Qr){\mathrm d}r ,
\label{SachsEl_FF}\\
G_M^p(Q^2)&=& \int_0^\infty \frac{2G_{(1/2)0}(r) F_{(1/2)0}(r)}{Q}\mu_Nj_1(Qr){\mathrm d} r.
\label{SachsMag_FF}
\end{eqnarray}
At zero momentum transfer, these form factors are normalized as $G_E^p(0)=e_p=1$, where $e_p$ is the proton  electric charge, and $G_M^p(0)=\mu_p $, with  $\mu_p$ standing for the magnetic dipole moment, whose experimental value is reported\- as\- $\mu_p=2.79284734462(82)\mu_N$ in \cite{PART}.

\noindent 
The goal of the present section is to test  as to what extent the ground state  Dirac spinor in (\ref{QSRD_spinor}) with the function in (\ref{F0})
is capable of capturing the essentials of the internal proton electromagnetic structure. In order to account for the property of the wave function to be defined over a finite interval, the Fourier transform has to be properly modified. Integral transforms of functions defined on  finite intervals on the real line have been studied for example  in \cite{Alonso}, where the modification of the plane wave has been obtained as
\begin{equation}
e^{i q r \cos \theta}\longrightarrow e^{i q R\chi \cos \theta},\quad \chi\in \left[0,\pi\right],
\label{Sherman}
\end{equation}  
amounting with the aid of (\ref{QSRD_spinor}), (\ref{F0}) to
\begin{eqnarray}
G_E^p(Q^2) &=& N_1\int _0^\pi  |{\widetilde F}_{00}^{(1/2)}(\chi)|^2j_0(QR\chi){\mathrm d}R\chi,  \label{GEP}\\
\frac{G_M^p(Q^2)}{\mu_N} &=& N_2\,\int _0^\pi |{\widetilde F}_{00}^{(1/2)}(\chi)|^2\frac{j_1(QR\chi)}{Q}{\mathrm d}R \chi ,
 \label{MEP}
\end{eqnarray}
where $N_1$ and $N_2$ are constants related to the spinor normalization accounting for  ~(\ref{trnsfrmd_wafuquasiradial_Dirac_spinor}). 

In so doing, the explicit expressions for the integrals to be evaluated below become
\begin{eqnarray}
G_E^p(Q^2)=N_1 \int_0^\pi \sin^{2(s+1)}\chi \exp \left(-\alpha_0 \chi \right)
j_0(RQ\chi){\mathrm d}R\chi,&&
\label{GEP_P}\\
\frac{G_M^p(Q^2)}{\mu_N}=N_2\int_0^\pi \sin^{2(s+1)}\chi \exp \left(-\alpha_0\chi \right)
\frac{j_1(R Q\chi)}{RQ}{\mathrm d}R\chi,
\label{GMP_P}\\
s=\sqrt{1-\alpha_s^2N_c^2}, \quad \alpha_0= 2\sqrt{1-\alpha_s^2N_c^2}\frac{mc^2R }{\hbar c}.
\label{constantes}
\end{eqnarray}
In the following we seek for values of  the three parameters: \,   $s$ (related to $\beta_0$ in (\ref{F0})), $R$, and $\alpha_0$ for which the experimental data on $G_E^p(Q^2)$, and $G_M^p(Q^2)$, together with their ratio, all taken from \cite{arrington} can be  adjusted by the  expressions in (\ref{GEP_P}), and (\ref{GMP_P}), respectively. \\

{}For that purpose we run a least mean square fit procedure over the entire data set including 
$G_E^p(Q^2)$, $G_M^p(Q^2)/\mu_p$, \- and \- $\mu_pG_E^p(Q^2)/G_M^p(Q^2)$ , finding the following parameter values:

\begin{equation}
\beta_0=-1.143789, \quad \frac{\alpha_0}{2}=2.60695, \quad R=1.10773\,\, \mbox{fm}.
\label{fitset}
\end{equation}

\begin{figure}
{\includegraphics[width=7.5cm]{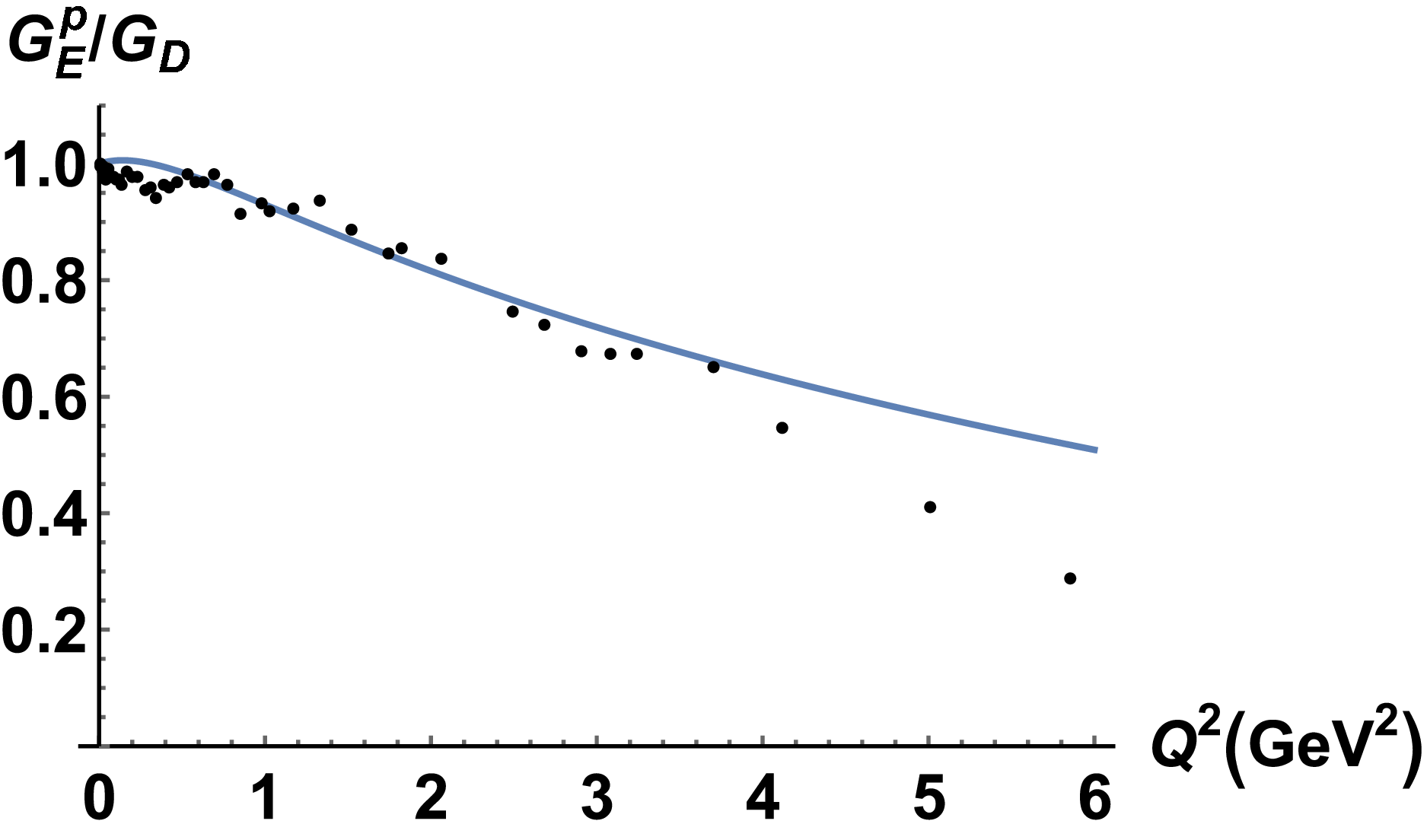}}
{\includegraphics[width=7.1cm]{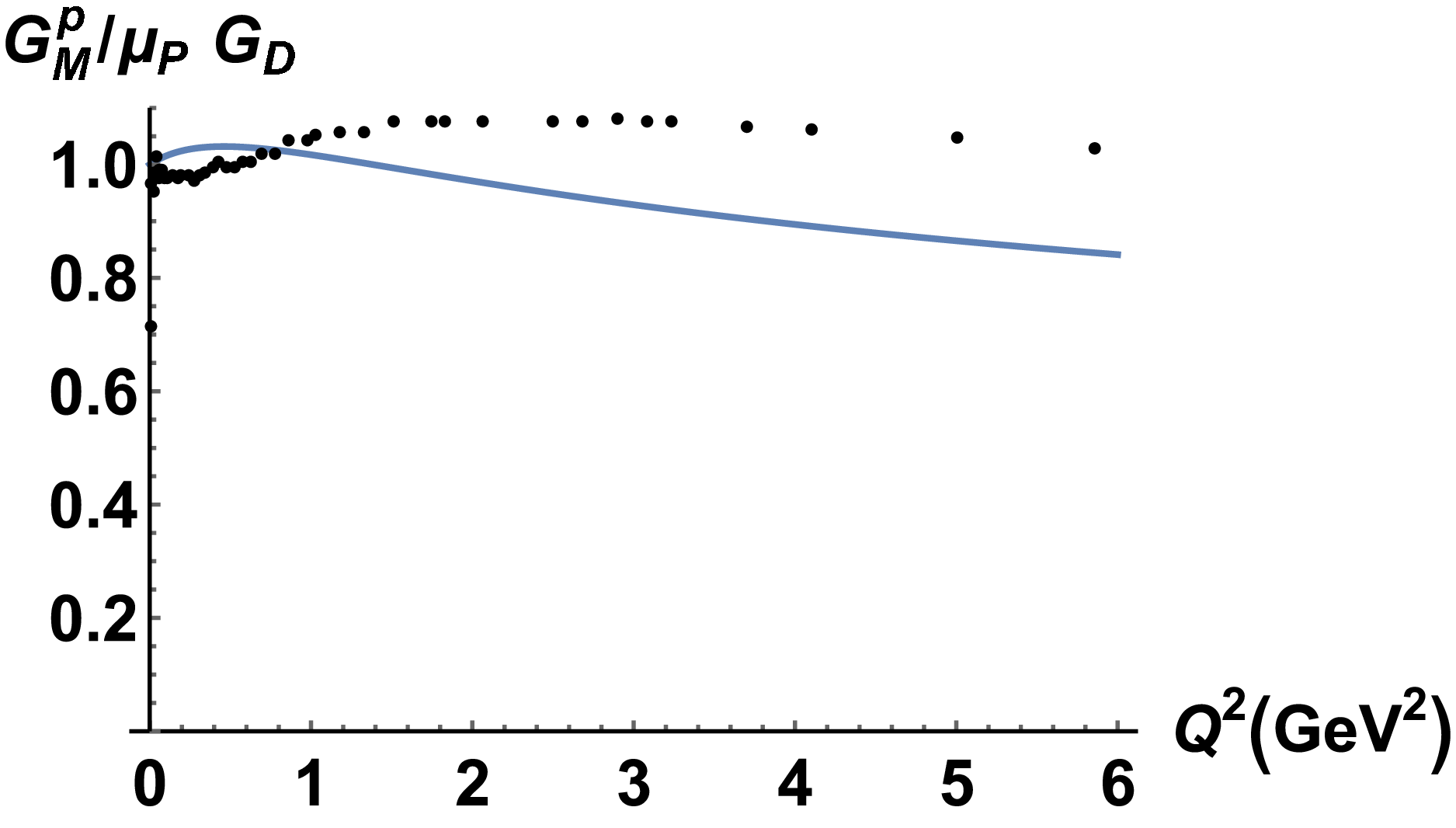}}
\caption{The ratios of the proton's electric charge $G_E^p$, and magnetic-dipole, $G_M^p/\mu_p$, form-factors with the dipole function, $G_D= \left(1+\frac{Q^2}{0.71}\right)^{-2}$, each normalized to one at origin, and calculated with 
the best fit parameter set  in (\ref{fitset}).
 Data taken from  \cite{arrington} and marked by points.} 
\label{analytic_expressions}
\end{figure}

\begin{figure}
\begin{center}
{\includegraphics{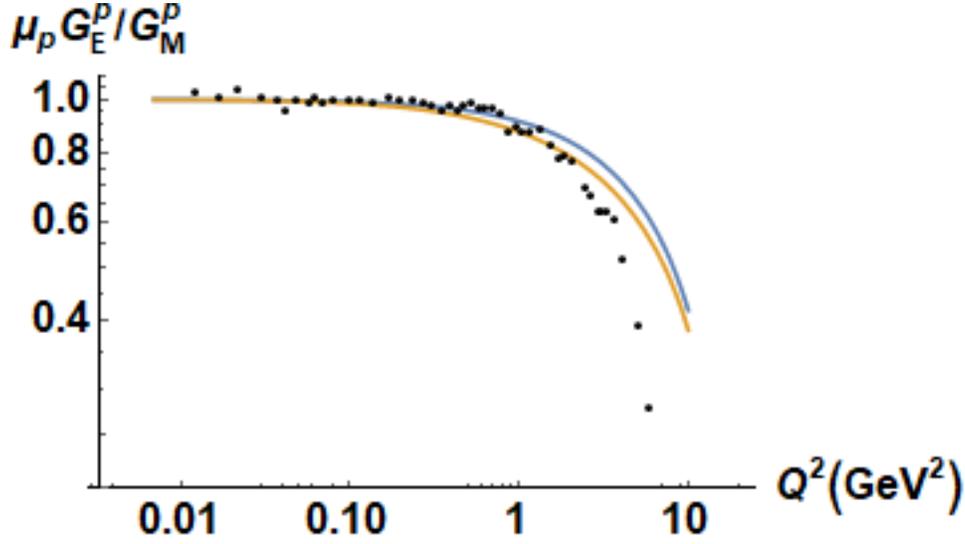}}
\end{center}
\caption{The $\mu_p G_E^p/G_M^p$ ratio for the numerically evaluated form factors displayed in the Figures   \ref{analytic_expressions} (yellow line) and for the analytic expressions in (\ref{GEp_anltc}) and (\ref{GMp_anltc}) (blue line) on a logarithmic
$Q^2$ scale of the horizontal axis.  Data taken from \cite{arrington} and marked by points.}
\label{ratio_plot}
\end{figure}

 The ratios of the  form-factors under investigation to the dipole function 
$G_D= \left(1+\frac{Q^2}{0.71}\right)^{-2}$ are plotted and compared to data in Fig.~ \ref{analytic_expressions}, while the
$\mu_pG_E^p/G_M^p$ ratio is displayed in Fig.~\ref{ratio_plot}.
The figures convincingly show  that the calculations realistically capture the proton's spin physics and the measured  shapes of the two form factors under consideration. Our results compare  in quality to the predictions of the light-front holographic formalism \cite{Mondal} as well as  other adequate models reviewed in \cite{Vanderhagen}.

\begin{quote}
It has to be stressed that the exponential fall-off of the  Dirac spinor wave function, brought about by the color confining dipole potential in (\ref{our_gpt}) that effectively accounts for confinement,  has been crucial for the satisfactory data description. {}For this reason we conclude that the potential in (\ref{our_gpt}) adequately accounts for the perception of  color  confinement by the proton's electromagnetic form-factors.
\end{quote}
 
\begin{figure}
\begin{center}
{\includegraphics[width=7.5cm]{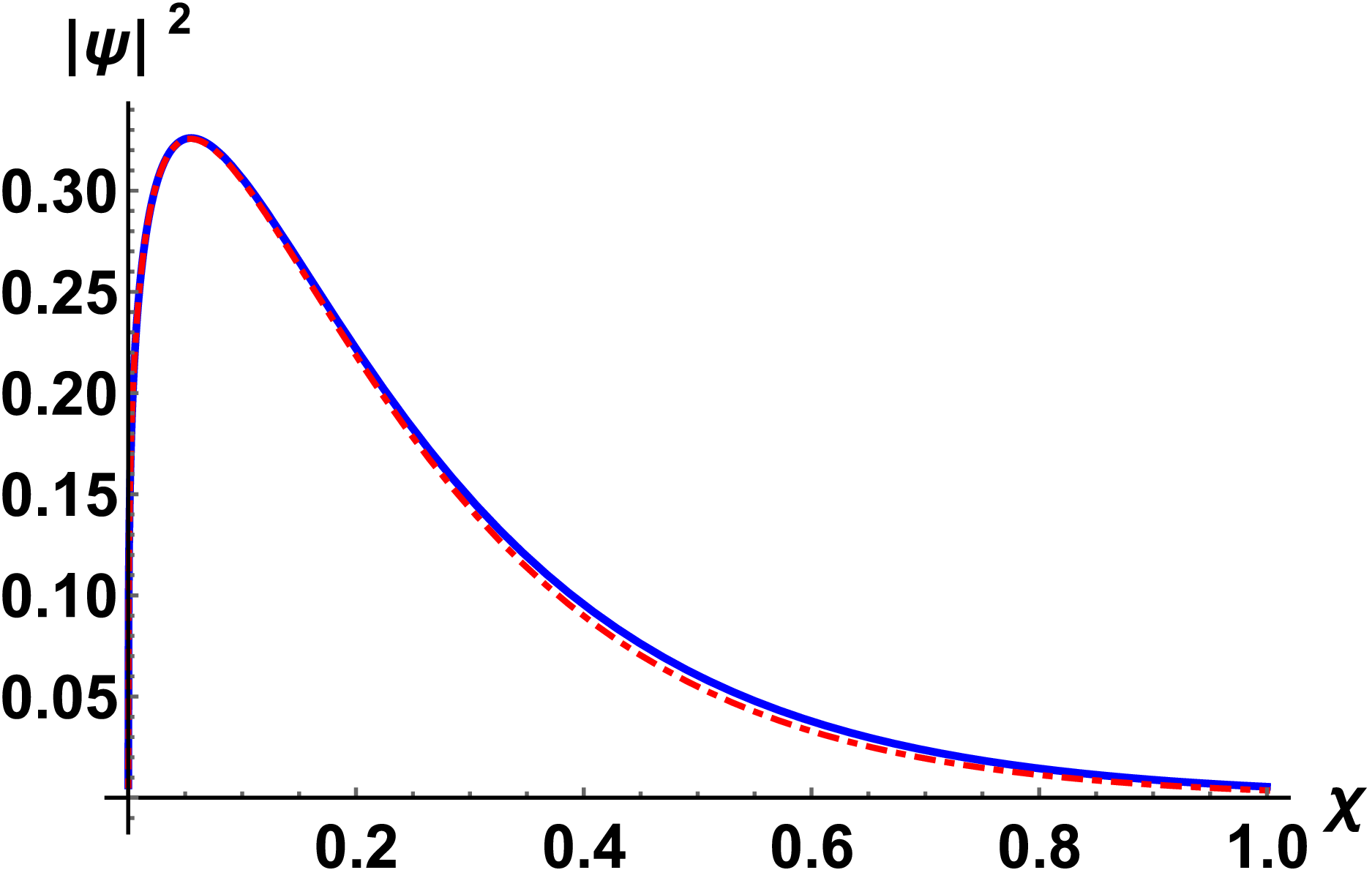}}
\end{center}
\caption{The shape of the probability density in the ground state, $|\Psi^{(\mbox{gst})}_{1/2,0}(\chi)|^2$, denoted in the figure by $|\Psi|^2$, and calculated for the parameters in (\ref{fitset}) once with  the exact wave function from eqs.~(\ref{QSRD_spinor})-(\ref{F0})  (solid line), and then in the  $\sin\chi \approx \chi$  approximation (dashed line).  } 
\label{dnsties}
\end{figure}

We furthermore  observe that as visualized by  Fig.~\ref{dnsties},  the probability densities calculated  once with the exact function ${\widetilde F}^{(1/2)}_{00}(\chi)$ in (\ref{F0}), and then  by its approximated form corresponding to,  $R\sin^{s+1}\chi\approx R\chi^{s+1}$, are practically coincident for the set of parameters in (\ref{fitset}). 
Thanks to this circumstance, our  ``quasi-radial'' Dirac equation effectively behaves as a flat-space radial equation though with a potential defined on a finite interval of the real line, a circumstance that in the hindsight justifies usage of the standard formulas in \cite{Kelley}.

\noindent
Moreover, extending in this very approximation the  integration in (\ref{GEP_P})-(\ref{GMP_P}) towards  the mathematically permitted infinity, practically does not alter the numerical results, but brings the advantage to allow one  to express the corresponding integrals, each normalized to one at origin, 
 in closed form,  which we  obtained by the symbolic software Matematica as,
\begin{eqnarray}
G_E^p(Q^2)&=& \frac{1}{(1+y^2)^{1+s}}\frac{\sin(2(1+s)\tan^{-1}y) }{2(1+s)y},\quad y=\frac{QR}{\hbar c \alpha_0},\label{GEp_anltc}
\end{eqnarray}
and

\begin{eqnarray}
\frac{G_M^p(Q^2)}{\mu_p}&=& \frac{1}{(1+y^2)^{1+s}}\frac{3}{2(1+s)(3+2s)y^2}\nonumber\\
&\times &\left( 
\frac{(1+y^2)^{\frac{1}{2}}}{(1+2s)y}
\sin\left(
(1+2s)\tan^{-1}y
\right) -
\cos \left( 2(1+s)\tan^{-1}y \right)
\right).
\label{GMp_anltc}
\end{eqnarray}
The comparison of the above expressions  to the numerical results is shown in Fig.~\ref{ratio_plot}. 
Finally, we also calculated the proton's charge and magnetic root mean-square radius and  magnetic moment \cite{Smith}, finding the following values,
\begin{eqnarray}
\sqrt{<r_p^2>_E}&=&\sqrt{< (R\chi)^2>} = 
\sqrt{
\frac{R^2(4+2s)(3+2s)}{\alpha_0^2}
}=0.79765 \,\, \mbox{fm},
\label{elm_cosnstasnts}
\end{eqnarray}

\begin{eqnarray}
\sqrt{<r_p^2>_M}&=& 
\sqrt{\frac{12R^2(2+s)(5+2s)}{10 \alpha_0^2}}=0.78358\,\, \mbox{fm},
\label{mag_constants}
\end{eqnarray}

\begin{eqnarray}
\frac{\mu_p}{\mu_N}&=&\frac{2}{3}\frac{M_pc^2}{\hbar c}\gamma <R\chi >\nonumber\\
&=&\frac{2}{3}\frac{M_pc^2\gamma R}{\hbar c} \frac{\Gamma (4+2s)}{2\alpha_0(1+s)\Gamma(2+2s)} =\frac{2}{3}\frac{M_pc^2\gamma R}{\hbar c}\frac{(3+2s)}{\alpha_0}=2.1910.
\label{mupvalue}
\end{eqnarray}
The proton  charge root mean square  radius lies by about $9$ \% below the data point of $0.84184(67)$  reported in \cite{Pohl}, while $\mu_p$ underestimates the well known experimental value \cite{PART}  of
$2.79284734462(82)\mu_N$  by about $20$\%. We explain the latter circumstance by the fact that we extracted the model parameters by  fitting  the $G_E^p$ and $G_M^p/\mu_p$ data and then calculated $\mu_p$ with these parameters according to the expression given in (\ref{mupvalue}). Stated differently, the $\mu_p$ value has not been explicitly included as an observable to be adjusted by the fit. Our predicted  proton  magnetic radius underestimates the data point of $0.86^{+0.02}_{-0.03}$ reported in \cite{Lorenz} by about 9\%. Nonetheless, in our opinion the results obtained are all pretty reasonable especially in view of the fact that they all have been evaluated by employing  same set of three parameters in (\ref{fitset}). The parameters $\gamma$, $\alpha_0$ and $s$ entering the expressions in (\ref{elm_cosnstasnts})-(\ref{mupvalue})  have been previously defined in the above equations (\ref{constants}), (\ref{parametri}), and (\ref{F0}). Their numerical values correspond to (\ref{fitset}).

A comment is in place on criticism regarding the frame-dependence of the  Sachs form-factors \cite{Miller}. Our model at the present stage is not formulated in a  Lorentz covariant fashion, though this is not a conceptual problem  but rather technical
issue. Indeed, as explained above in (and around) the equation (\ref{effctv_Mnk}), the $S^3$ Laplacian   can be transformed to the effective Minkowski metric as explained around and in the equation  (\ref{effctv_Mnk}). On such a plane Minkowski space-(conformal)time  one can define Lorentz transformations, four vectors, covariant couplings to the electromagnetic field, and even switch to light-front variables, a project for future research, whose mentioning here has the purpose to point out that Lorentz invariance must  not be  beyond reach of models formulated on closed spaces.  Surprisingly, the lack of manifest Lorentz covariance of the model under investigation seems to have a minor effect on its predictions.
Indeed,  we extracted the Dirac form factors $F_1^p(Q^2)$ and $F_2^p(Q^2)$  from $G_E^p(Q^2)$ and $G_M^p(Q^2)$ as
\begin{eqnarray}
F_1^p(Q^2)&=&\frac{G_E^p(Q^2) +\frac{Q^2}{4M^2}G_M^p(Q^2)}{1 +\frac{Q^2}{4M^2}},\label{F1}\\
\kappa F_2^p(Q^2)&=&\frac{G_M^p(Q^2) -G_E^p(Q^2)}{1 +\frac{Q^2}{4M^2}},\label{F2}
\end{eqnarray}  
using our  calculated $\kappa =1.191$ value and the experimental proton mass, and compared to the data set reported in \cite{Cates}, also used in \cite{Mondal} 
in the evaluation of hadron electromagnetic form-factors within the LF-HQCD theory. The results are shown in Figs.~5,6. No significant discrepancy between predictions and data is observed  for the Pauli form factor $F_2$.  However one expects the Dirac form-factor, $F_1$, whose $Q^2$ dependence is more complicated than that of $F_2$,
to be more affected by Lorentz boost effects. Indeed, in $F_1$  one observes a small but detectable underestimation of data within the region between $0.5$ GeV$^2$ to $2$ GeV$^2$ where Lorentz effects are expected to be viable. We interpret this minor deviation from data as an artifact of the missing Lorentz invariance of our method in its present form. However, at the same time, we think that the smallness of the effect and its gradual disappearance at higher  $Q^2$ is due to the favorable r\'ole played by the exponential fall off of the Dirac wave functions which above 2 GeV$^2$ seem to take over Lorentz boost corrections. In conclusion, we
 interpret the satisfactory description of the proton electromagnetic form factors within our framework  as a supremacy  over the kinematic Lorentz invariance
of the dynamical conformal symmetry, implemented by the Dirac equation of the present method  in a way similar to the LF-HQCD Dirac equation.

\begin{figure}
{\includegraphics[width=7.5cm]{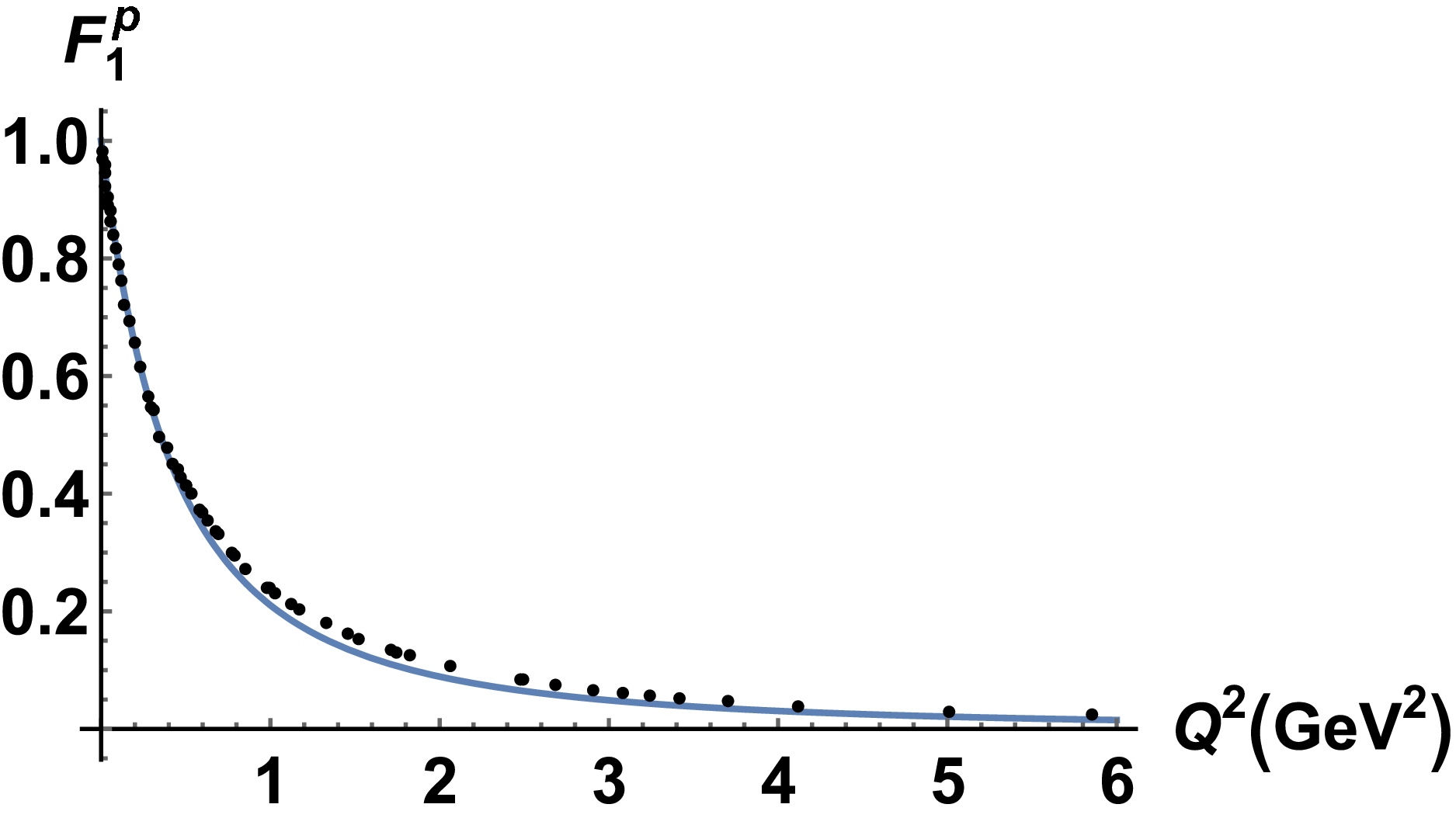}}
{\includegraphics[width=7.1cm]{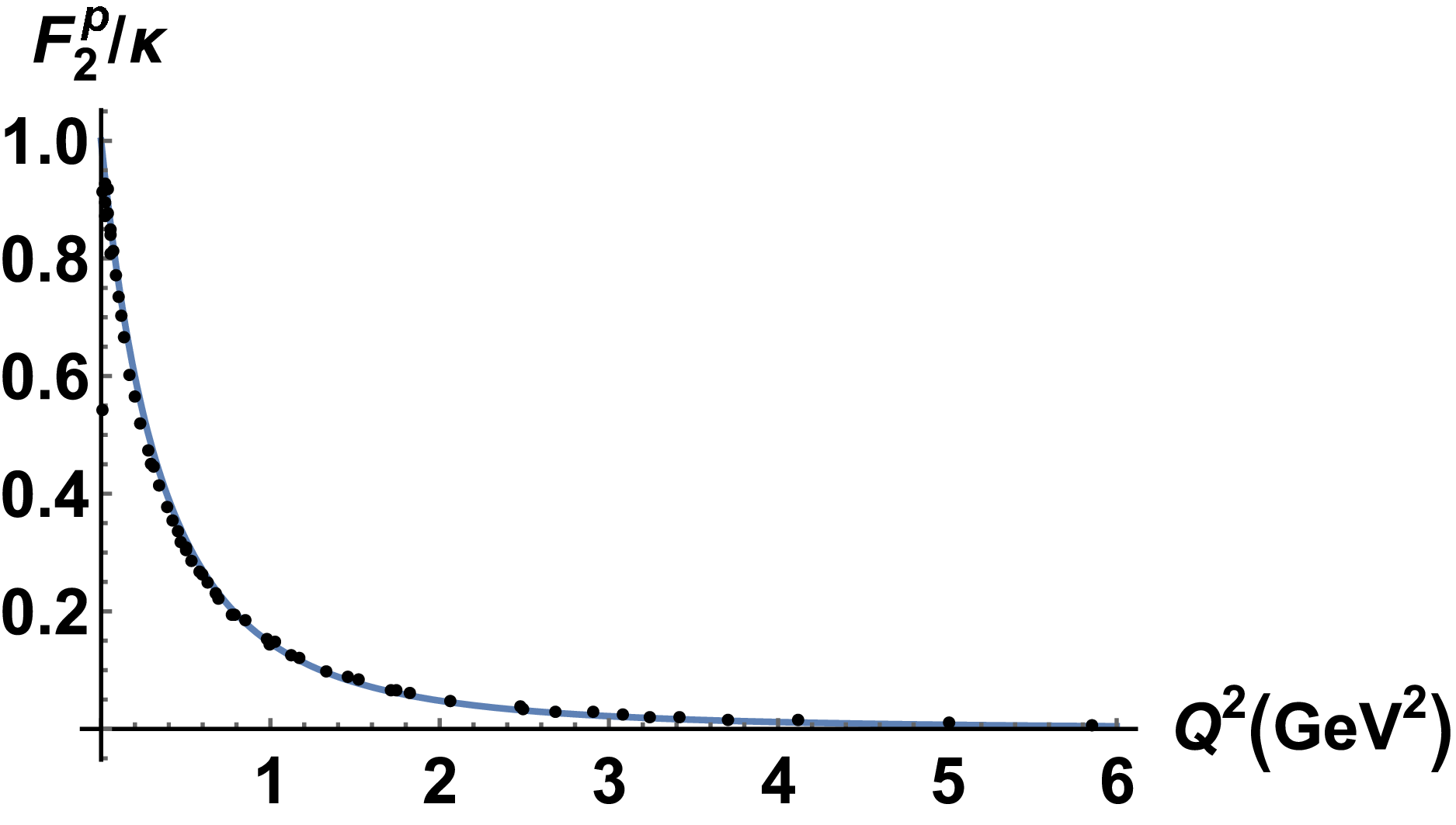}}
\caption{Dirac's  $F_1^p$ (left)  and $F_2^p$ (right)  form factors.
 Data taken from  \cite{Mondal} and marked by points.} 
\label{DiracFFS}
\end{figure}

\begin{figure}
\begin{center}
{\includegraphics[width=7.5cm]{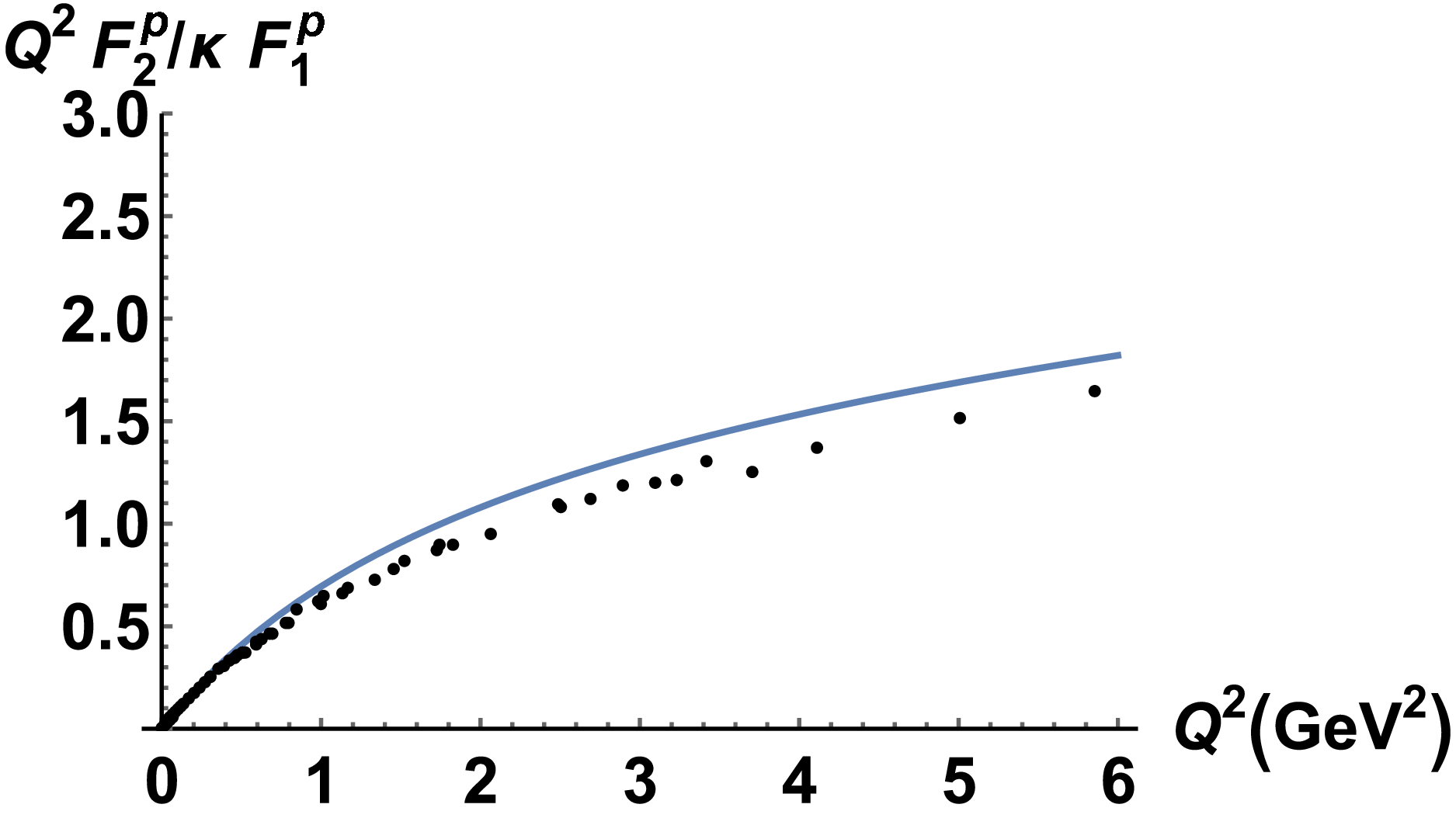}}
\end{center}
\caption{The $Q^2 F_2^p/F_1^p$ ratio.  Data taken from \cite{Mondal} and marked by points.}
\label{ratioF2F1}
\end{figure}

As one more test of the credibility of the here advocated model we check below  compatibility of our predicted  $\mu_n/\mu_p$ ratio with data. The most immediate approach to the neutron magnetic dipole moment, $\mu_n$, in units of $\mu_N$,   within the present framework is found through  its electric charge form factor and its relationship to  the respective proton form factor  while making use of a Galster inspired parametrization   \cite{Gentile}, here chosen as

\begin{equation}
G_E^n(Q^2)=-\frac{\mu_n\tau}{1+B\tau}G_E^p(Q^2), \quad \tau=\frac{Q^2}{4M_n^2}.
\label{Galster}
\end{equation}

In adjusting by the $\mu_n$ and $B$ parameters the expression in (\ref{Galster}) to the data  taken from \cite{Gentile},
the  $\mu_n$-value emerges as

\begin{equation}
\mu_n=-1.58661\mu_N,
\label{Muntr}
\end{equation}
and similarly to the proton's magnetic dipole moment, underestimates the data point given by, $-1.91304245(45)\mu_N$,  by about 20\%. The resulting   
$\mu_n/\mu_P$ ratio,
\begin{equation}
 \frac{\mu_n}{\mu_p} = -0.7244,
\label{munmp}
\end{equation} 
overestimates the experimental data point of, $-0.6849\-7934\-(16)$,  by only few percents.
The neutron's electric charge form-factor is displayed in Fig.~7.

\begin{figure}
\begin{center}
{\includegraphics[width=7.5cm]{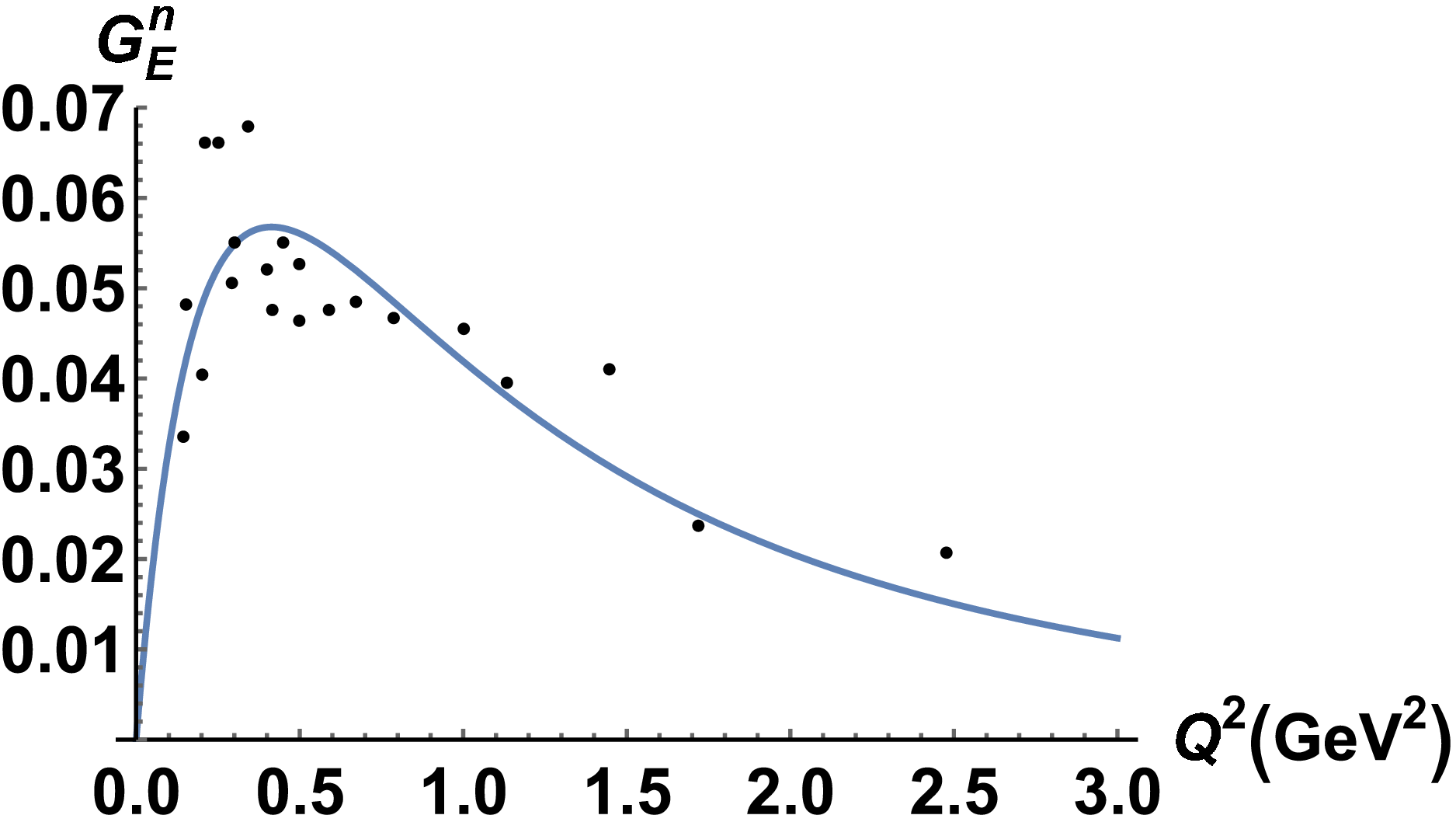}}
\end{center}
\caption{The electric charge form factor of the neutron following from (\ref{Galster}) (solid line) for $B=2.54061$ and $\mu_n=-1.586\mu_N$ in comparison to data taken from \cite{Gentile} (points). The accuracy of the fit is illustrated by the value of $\sigma_E^n = \sqrt{\sum_{i=1}^{i=N}
\left(\left[G_E^n(Q^2_i)\right]^{th} - \left[ G_E^n(Q^2_i)\right] ^\ast
\right)^2}/ (N-N_P) = 0.00789728$,
where asterisks denotes the data points, $N=22$ is the number of data points, while  $ N_P=2$ refers to the number of parameters.}
\label{GEM}
\end{figure}

{}Finally, our approach also allows for a convenient parametrization of the neutron magnetic form factor, $G_M^n(Q^2)$ in so far as  when normalized to one at origin, $G_M^n(Q^2)/\mu_n$,  it can be pretty well approximated by the normalized proton magnetic 
form factor, $G_M^p(Q^2)/\mu_p$. In this way,  equality of the proton and neutron magnetic radii is predicted. Such an approximation is compatible with 
the fact that the neutron's magnetic radius, $r_M^n=0.88\pm 0.05$fm \cite{Lorenz}, is by about only $2$\% larger than the proton's magnetic radius.    
The comparison of $G_M^n(Q^2)=\mu_nG_M^p(Q^2)/\mu_p$ to data is displayed in Fig.~8, and presents itself  pretty convenient, indeed. 
\begin{figure}
\begin{center}
{\includegraphics[width=7.5cm]{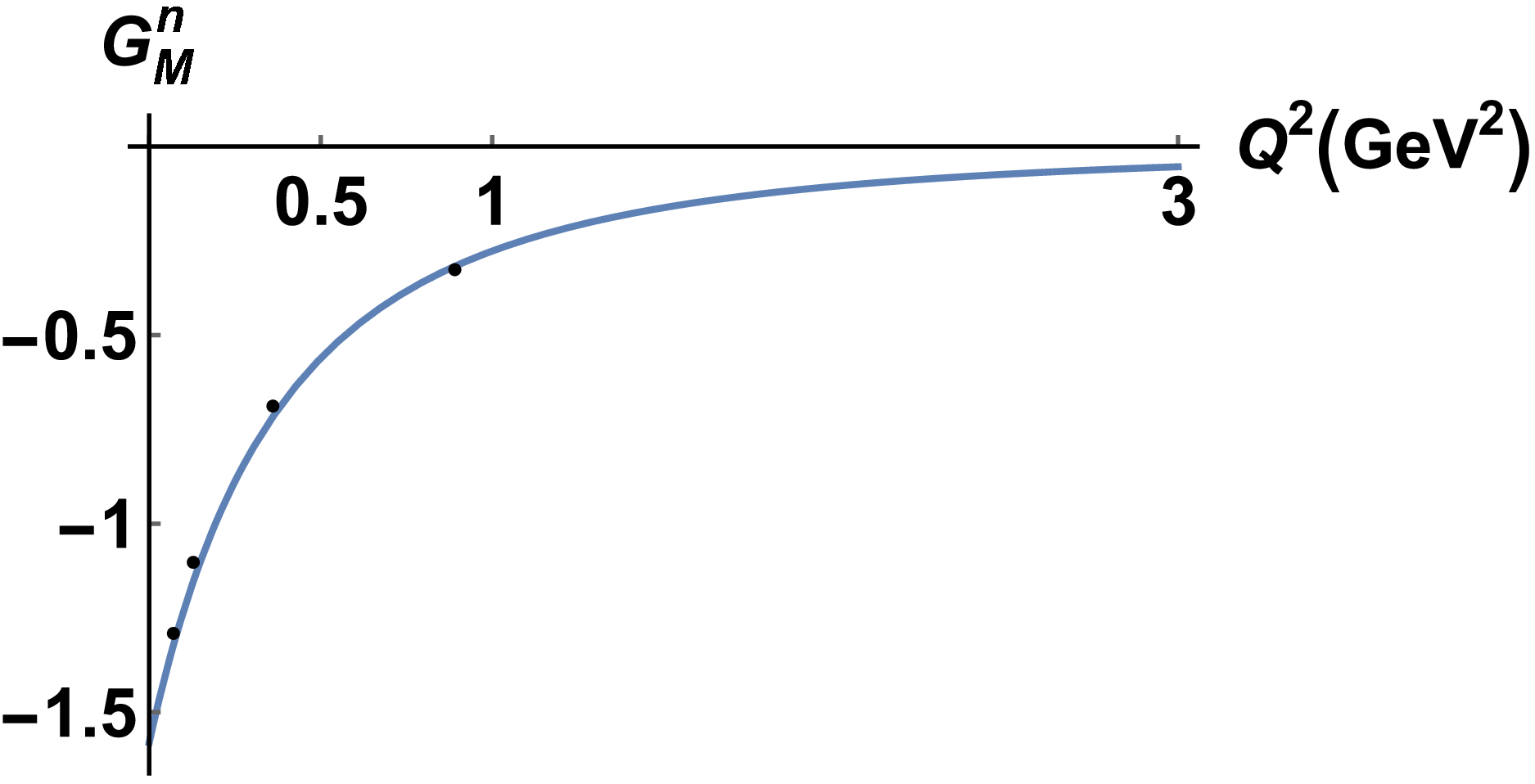}}
\end{center}
\caption{The neutron magnetic form factor, $G_M^n(Q^2)$, here denoted by $G_M^n$, and calculated as $G_M^n(Q^2)=\mu_nG_M^p(Q^2)/\mu_p$ with $\mu_n$ and $G_M^p(Q^2)$ from the respective eqs.~(\ref{Muntr}), and (\ref{GMp_anltc}), and with the same parameter set in (\ref{fitset}). As trim points representative of  the experimental data set we have chosen the four data points reported in the Table 2  of ref.~\cite{Kubon}. There, the measurements have been preformed  at $Q^2$ values of $Q^2=0.071$, $0.125$, $0.359$, and $0.894$ GeV$^2$, and the $G_M^n/(\mu_n G_D)$ values (with $G_D$ as always standing for the dipole function already given in the caption to Fig.~2) have been extracted as $0.990\pm 0.013$, $0.967\pm 0.013$, $0.989\pm 0.014$, and $1.062 \pm 0.014$, respectively. The figure shows that our parametrization captures well the tendency of data modulo that data extrapolate at origin at $\mu_n=-1.91304245(45)\mu_N  $ while in our calculation the absolute value of this point appears by about 20\% lower the experimental one.}
\label{GnM}
\end{figure}

\begin{quote}
Recapitulating, the framework presented here provides a reasonable description of  the proton's electric-charge and magnetic-dipole form factors
together with their ratio. Also the neutron's charge and magnetic form factors came out reasonably, amounting to a  $\mu_n/\mu_p$ ratio in a good agreement with data. Furthermore, the proton's electric-charge and  magnetic radii have been recovered within an accuracy of few percents. As a result, a variety of observables could be  predicted  within acceptable range of accuracy by the aid of only three parameters. Such has been possible, in our opinion,  by the virtue of the conformal symmetry of the non-power quark potential defining the Dirac spinors of the proton.    

\end{quote}

Before closing the current section, we like to recall on the relationships of the model parameters 
to fundamental constants in QCD. As repeatedly stressed through the text, the magnitude of the potential used was determined by the product of  the strong coupling $\alpha_s$ and the numbers of colors, $N_c$ in QCD. In adopting the established $N_c=3$ value, the partner potentials in (\ref{H1})-(\ref{Pot_H2})  depend on the  parameter $\alpha_s$, absorbed by  $\beta_0$  in (\ref{F0}) in combination with (\ref{parametri}), and, via $E_{j\ell }$ in (\ref{gst_energy}), on  the mass parameter $mc^2$, absorbed by $\alpha_0$ in (\ref{F0}). 
In being less than the nucleon mass, this $mc^2$ parameter comes out pretty reasonable. However, so far we have not elaborated any approach to the extraction of the nucleon mass from the reduced mass.
Therefore, the fit by the two parameters,  $\beta_0$, and $\alpha_0$  in (\ref{fitset}) can be converted to  a fit by $\alpha_s$, and a  mass. The Table I shows these values. We interpret the small constant value for $\alpha_s$ as an average of the running coupling, $\alpha_s(Q^2)$, over the $0 \leq Q^2 < 6\, \mbox{MeV}^2$ range of evaluation of the form-factors under investigation. \\

\noindent
It is perhaps also interesting to notice that applying the present formalism to  the H Atom, i.e. placing it  on  a closed $S^3$ space, predicts a hyper-radius value of the order of $10^{-3}$ cm and thereby by eight orders of magnitude larger than the $H$ Atom size, a result concluded from fitting magnetic dipole matrix elements to hydrogen hyper-fine structure effects   \cite{Bessis2}. In contrast,   the hyper-radius of the strong space obtained here  is of the order of $10^{-13}$ cm and comparable with the nucleon size. Stated differently, electromagnetic processes show a clear  preference towards  a plane space-time, where standard electrodynamics  can be applied in the evaluation of the physical properties of the electron, such as its gyromagnetic factor.

\begin{table} 
 \begin{center}
\begin{tabular}{|c|cccc|}  
\hline 
form factors  & $\alpha_s$  &  $mc^2$  [MeV]  & $R$ [fm] & $\sigma^p_{E/M}$ \\ 
\hline 
$G_E^p$  & 0.3298 & 469.27  & 1.10773  &  0.0135263 \\
\hline 
$G_M^p$  & 0.3298 & 469.27 & 1.10773   & 0.0461983\\
\hline
\end{tabular} 
\caption{
Values of the strong coupling, $\alpha_s$, and the mass $mc^2$ parameter (first and second columns) extracted from fitting proton's electric-charge and magnetic-dipole from factors together  with their ratio  by the parameters $\beta_0$ and $\alpha_0$  in (\ref{fitset}). The  third column contains  the values of the length scale $R$ from same parameter set (\ref{fitset}). In the last column the $\sigma_{E/M}$ value,illustrative of  the quality of our date fit, is defined as,
$\sigma^p_{E/M}=\sqrt{\sum_{i=1}^{i=N} \left(\left[ G_{E/M}^p(Q^2_i)\right]^{th} - \left[ G_{E/M}^p(Q^2_i)\right]^\ast\right)^2}/(N-N_P) $, with the asterisks  denoting the experimental data points whose number is $N=47$, while $N_p=3$ stands for the number of the three recurring parameters. 
}
\end{center}
\label{Table3} 
\end{table}

\section{Summary and conclusions}
The goal of the present work has been to study influence of dynamical conformal symmetry and color confinement in the infrared on proton's electromagnetic form factors through employing in a Dirac equation a non-power (trigonometric) potential with these  properties, which has earlier been shown in \cite{EPJA16},\cite{Addendum_EPJA} to be suited for description of meson spectra. The potential is given in the above eq.~(\ref{our_gpt}),  and allowed one to  obtain  spinor-wave functions  adequate for the evaluation of the observables under investigation.  The Dirac equation with this interaction has been formulated and  approximately solved by the tools of the super-symmetric quantum mechanics. The solutions have then  been employed in the evaluation of the electric-charge and magnetic-dipole  form factors of the proton,  their ratio, the electric-charge and magnetic-dipole form factors of the neutron, the root mean square charge and magnetic proton radii, the proton and neutron magnetic moments and their ratio,  finding not only pretty good data description by the same set of parameters in (\ref{fitset}), but also quite  reasonable approximations to the numerical results by the closed form expressions, given  among others in (\ref{GEp_anltc})-(\ref{GMp_anltc}), and (\ref{elm_cosnstasnts})-(\ref{mupvalue}). Our analyzes revealed the notable role played specifically  by the color confinement, that ensured a satisfactory data description mainly in  providing  the exponential fall-off of the Dirac spinor wave function in (\ref{F0}) by virtue of the color confining and conformal dipole potential in (\ref{our_gpt}). We conclude that dynamical  conformal symmetry and color confinement in the infrared are compatible with data on the electric-charge and the  magnetic-dipole form factors of the proton, as well as with the deviation of their ratio from the dipole scaling rule. 
 We hope that the present study could convincingly demonstrate utility of non-power potentials in quark models.

\end{document}